\documentclass[seceq,supplement]{ptptex}

\usepackage{graphicx}

\title{
Hartree-Fock+BCS Approach to Unstable Nuclei with the Skyrme Force
}

\author{
Naoki {\sc Tajima}\footnote{
E-mail address: tajima@quantum.apphy.fukui-u.ac.jp}
}

\inst{
Department of Applied Physics, Fukui University,
Fukui 910-8507, Japan
}

\recdate{
}

\abst{
We reanalyze the results of our extensive Hartree-Fock + BCS
calculation from new points of view paying attention to the
properties of unstable nuclei.
The calculation has been done with the Skyrme SIII force for
the ground and shape isomeric states of 1029 even-even nuclei
ranging 2 $\le Z \le$ 114.
We also discuss the advantages of the employed three-dimensional
Cartesian-mesh representation, especially on its
remarkably high precision with apparently coarse meshes
when applied to atomic nuclei.
In Appendices we give the coefficients of finite-point numerical
differentiation and integration formulae suitable for Cartesian mesh
representation and elucidate the features of each formula and the
differences from a method based on the Fourier transformation.
}

\newcommand{\mbf}[1]{\boldsymbol{#1}}

\begin{document}


\maketitle

\section{Introduction}

Among theoretical attempts which aim at treating all the nuclides in a
single framework, the simplest category seems to be the mass formula.
Hence, let us focus on the nuclear masses in the first
place, although the scope of this paper is not restricted to the
masses.  The most familiar mass formula is certainly that of Bethe and
Weizs{\"a}cker\cite{We35,BB36}, which expresses the nuclear masses as
a function of the number of neutrons ($N$) and that of protons ($Z$)
as,
\begin{equation} \label{e_BW_mass_formula}
  E(N,Z) = a_{\rm V}         A      + a_{\rm S}     A^{2/3}
         + a_{\rm I} (N-Z)^2 A^{-1} + a_{\rm C} Z^2 A^{-1/3},
\end{equation}
where $A=N+Z$ is the mass number.  The four terms on the right-hand
side of this equation are called respectively the volume, the surface,
the symmetry and the Coulomb terms.  These terms come from the
liquid-\-drop picture of the atomic nucleus, in which the density
distribution of nucleons is assumed to be spatially uniform inside
nuclei and the uniform density value is common to all the nuclei.

Incidentally, it is worth stressing that even the form of the symmetry
term is completely understandable within the liquid-\-drop picture.
It is not necessary to bring up the Fermi gap model like, e.g.,
in p.~5 of Ref.~\cite{RS80}.
The form of this term can be naturally derived for a liquid drop
by admitting that the coefficient of the volume term should
have a dependence on the ratio of the constituents.
Because one knows that the volume energy becomes maximum when the
proportion of protons (or neutrons) is 50\%, the volume term should be
modified as,
\begin{equation}
a_{\rm V}A \; \longrightarrow\;
\left\{ a_{\rm V} + a_{\rm V}' \left( \frac{1}{2}- \frac{Z}{A} \right)^{2}
\right\} A
\; = \; a_{\rm V} A + \frac{a_{\rm V}'(N-Z)^2}{4A},
\end{equation}
if one retains terms only up to the second order in $Z/A$.
On the right-hand side of this equation, the first term is the
original volume term and the second term is identical to the symmetry
term by rewriting $\frac{1}{4}a_{\rm V}' = a_{\rm I}$. If one applies
a similar consideration to the surface term one obtains a so-\-called
surface symmetry term, $a'_{\rm I} (N-Z)^2 A^{-4/3}$. This term is
sometimes added to the liquid-\-drop mass formula of
Eq.~(\ref{e_BW_mass_formula}) recently.

Now, by changing the values of the four coefficients
($a_{\rm V}$, $a_{\rm S}$, $a_{\rm I}$ and $a_{\rm C}$),
the root-mean-square (r.m.s.) error from all the experimental data of
about 2,000 even-even nuclei available at present\cite{AW93,AW95} can be
decreased down to 3.5 MeV.  Inclusion of the surface symmetry
term can decrease the error further.  The accuracy of 3.5 MeV is
very small compared with the binding energies of heavy nuclei ($\sim$
1 GeV).  However, the requirement for theoretical predictions of the
energies of unknown nuclei is something more precise.
The separation energy of a neutron, $S_{\rm n}$, decreases as the
neutron number $N$ is increased.
Its changing rate can be roughly estimated to be
\begin{equation} \label{e_Sn}
\frac{\partial S_{\rm n}}{\partial N} \; \simeq \;
- \frac{20}{\left( N^2 A \right)^{1/3}} \; (\mbox{MeV})
\; = \; - \frac{26}{A} \; (\mbox{MeV},
\;\;\mbox{\small for $N=\frac{2}{3}A$}),
\end{equation}
of which the right-hand side of the first (approximate) equality
is the reciprocal of the single-particle
level density at the Fermi level obtained by assuming a harmonic
oscillator potential with $\hbar \omega_{\rm osc} = 41 A^{-1/3}$ MeV.
The right-hand side of the last equality,
which is the expression for nuclei on the neutron drip
line, becomes as small as $100$ keV for the heaviest nuclei.
It means that
the precision of mass predictions must be 100 keV in order to
predict the location of the neutron drip line.

The first step to decrease the error is to take into account the shell
effect. For example, the TUYY mass formula\cite{TUY88} achieved an
r.m.s.\ mass error of 538 keV.  The number of fitting
parameters are 6 for the gross part (corresponding to the parameters
of the Bethe-Weizs{\"a}cker formula) while that for the shell part is
as many as 269: There is one parameter for each value of $Z$ in an
interval $1 \le Z \le 112$ and one for each value of $N$ in $1 \le N
\le 157$.  Generally speaking, less number of parameters are
preferable for the reliability of the extrapolation to nuclei not
synthesized yet.  One usually switches to less phenomenological models
in order to reduce the number of parameters.

What should be considered next is that the shell effect is strongly
dependent on deformation.
Indeed, the research group which presented the TUYY mass formula has
started the development of a new method which includes
the concept of
deformation \cite{KY00,KUT00}.
Of course there are different directions to proceed. For example,
a model based on the shell-model configuration mixing
has achieved an r.m.s.\ error of 375 keV with
28 parameters\cite{DZ95}.
The latest good review on the methods of theoretical predictions of
nuclear masses can be found in Ref.~\cite{Uno00}.

The most elaborately developed model which takes into account
deformation seems to be the finite-range droplet model with a
microscopic shell correction (FRDM), whose latest update was done by
M{\"o}ller et al.\ \cite{MNM95} \hspace*{5mm}
Another extensive calculation was carried out by
Aboussir et al.\ \cite{APD92} in the extended Thomas-Fermi plus
Strutinsky integral method (ETFSI).

These two methods can be regarded as approximations to the
Hartree-Fock (HF) method.  The straight-forward solutions of the HF
equations including deformation require long computation time even
with present computers in order to cover thousands of nuclei.
We understand that the first calculation of such a large scale which
have been published is our calculation\cite{TTO96}
done in the framework of
the HF+BCS method with the Skyrme SIII force.
After the publication of our paper\cite{TTO96},
only a few similar extensive calculations have been done.
Tondeur et al.\ have performed a set of such extensive calculations in
the framework of the HF+BCS and proposed a new Skyrme force parameter
set MSk7 best suited for their truncated oscillator
basis \cite{TGP00,GTP01}.
Another attempt have been carried out in the relativistic mean-field
framework \cite{SHT96}.
%
%

In section 2, we discuss about the Cartesian mesh
representation of the wave functions, which is a feature of our
calculations. Details not included in Ref.~\cite{TTO96}
are relegated to the Apppendices.

In section 3, we reanalyze the results of our extensive
calculation in order to give discussions from new points of view and
to compare the results with those of other models.

%
%


\section{Cartesian-Mesh Representation}

The most usual method to solve the mean-field equations without assuming
the spherical symmetry is to expand single-particle wave functions in
terms of truncated anisotropic harmonic-oscillator
basis \cite{Va73,DG80}.
Instead, we employ a three-di\-men\-sional Cartesian-mesh
representation \cite{BFH85}:
We put a nucleus in a box containing $\sim 10^4$ mesh points and
express each single-particle wave function in terms of its values at
the mesh points.
The advantages of this representation can be summarized as follows:

(1) It has no prejudice concerning the shape of the nucleus. On the
other hand, in methods based on expansions in the anisotropic
oscillator basis, the shape of the solution should be similar to that
of the basis in order that the truncation of the basis does not affect
the solution. To fulfill this requirement, one usually solves the
mean-field equation imposing constraints on the quadrupole moments
such that the resulting shape agrees with the anisotropy of the basis.
Then, to obtain the ground state, one has to optimize the anisotropy
of the basis such that the energy of the solution is minimized. This
procedure is not only cumbersome but also unmanageable in treating
exotic states: For example, if the protons and the neutrons may have
different shapes, the dimension of the parameter space for the
optimization is squared.  On the other hand, in the mesh
representation, one can treat various shapes with the same mesh.
Consequently, the optimization procedure is not necessary.

(2) In the mesh representation the asymptotic form of the wave
function at large radius can have arbitrary form, while in the
oscillator basis representation it must have a Gaussian tail and
therefore loosely bound single-particle wave functions
like neutron halo\cite{Tan85} cannot be correctly described.

\begin{figure}
\centerline{\includegraphics[angle=-90, width=125mm]{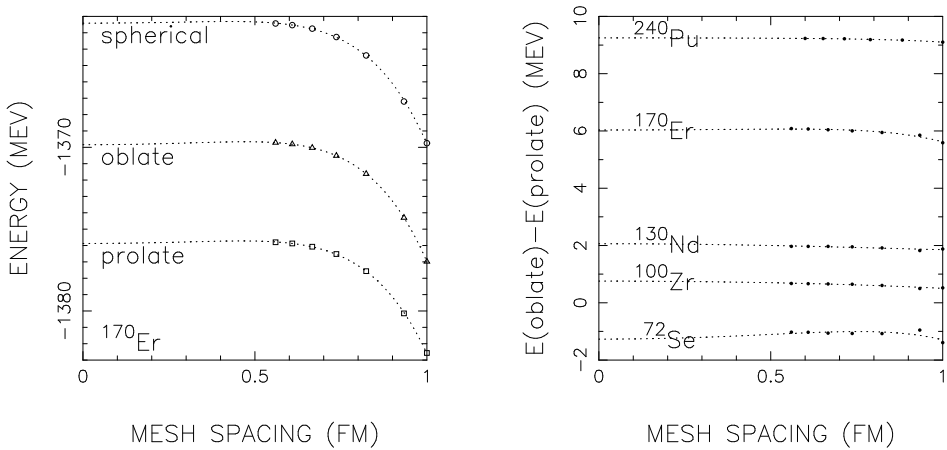}}
\caption{
Left-hand portion shows
the total energy of $^{170}$Er (for the prolate minimum, the
oblate minimum, and the spherical saddle point)
versus mesh spacing $a$. The circles,
triangles, and squares are the results of calculations while the
curves represent functions $E=c_0 + c_2 a^2 - c_6 a^6$ where $c_i$ are
determined through least-square fittings.
Right-hand portion gives the energy difference between prolate and
oblate minima for five nuclei.
} \label{g_masprec}
\end{figure}

(3) Among systems of many particles found in nature,
the atomic nucleus is a very suitable object to apply the
Cartesian-mesh representation: Inside nuclei, the density is roughly
constant and therefore the local Fermi momentum is also constant
everywhere. This situation favors an expansion in the plane wave
basis, to which the Cartesian mesh representation can be an accurate
approximation by using appropriate approximation formulae for
derivatives and integrals as discussed in Appendices.
Consequently, apparently coarse meshes can result in unexpectedly high
precision: With mesh spacing $a \sim$ 1 fm, there are only a few mesh
points in the surface region. Nevertheless,
the relative error of the total energy and the quadrupole moment with
this mesh spacing is as small as 0.5 \% for heavy nuclei.
The left-hand portion of Fig.~\ref{g_masprec}
shows how the total energy of $^{170}$Er changes as a function of the
mesh spacing. At $a$=1 fm, the error is $\sim$7 MeV, which is indeed
small (0.5\%).
Furthermore,
as shown in the right-hand portion of the figure,
the energy difference between prolate and oblate
solutions is by far more precisely determined (error $\sim$ 0.4 MeV)
with $a$=1 fm.
Owing to this independence of the error from the shape, we have found
it possible to correct the total energy by adding a simple function of
only $Z$ and $N$ to decrease the error down to $\sim$0.2 MeV

(4) The algorithm of the calculation is simple.
This simpleness increases the reliability of computer programs based on
the algorithm.
It also makes the programs efficiently executable with vector-processor
type computers.
The programs can also be fitted to each unit of massively parallel
computers because the necessary memory is not huge.


\section{HF+BCS Calculations with the Skyrme Force}

Using the Cartesian mesh representation, we have calculated the ground
and shape isomeric states of even-even nuclei with
$2 \leq Z \leq 114$ and $N$ ranging from outside the proton drip line
to beyond the experimental frontier in the neutron-rich side.
We have obtained spatially localized solutions for 1029 nuclei, together
with the second minima for 758 nuclei.
Details not described in this paper can be found in
Refs.~\cite{TTO96,TOT94,TTO98}.

The Skyrme force\cite{Sk56,VB72} is an effective interaction widely
used in mean-field calculations. It is essentially a zero-range force
but modified with the lowest order momentum dependences to simulate
the finite-range effects, a density dependence to reproduce the
saturation, and a spin-orbit coupling term.
The SIII\cite{BFG75} is one of the many parameter sets proposed for
the Skyrme force.
It features good single-particle spectra and accurate $N-Z$ dependence
of the binding energy\cite{TBF93}.

We have used a computer program named {\sl EV8}\cite{BFH85}.
In the program, one places an octant of a nucleus in a corner
of a box (13 $\times$ 13 $\times$ 14 fm$^3$ for $Z \le 82$ and
14 $\times$ 14 $\times$ 15 fm$^3$ for $Z > 82$),
imposing a symmetry with
respect to reflections in $x$-$y$, $y$-$z$, and $z$-$x$ planes (the
point group $D_{2h}$\cite{Ham62}).
The mesh spacing is 1 fm as explained in the last section.

For the pairing, we employ a seniority force, whose pair-scattering
matrix elements are defined as a constant $G_{\tau}$
($\tau$ distinguishes between proton and neutron)
multiplied by a cut-off factor which is a function of the
single-particle energy $\epsilon$.  The cut-off is set at $\epsilon$ =
(Fermi level + 5 MeV) with smearing width of 0.5 MeV.  For neutrons,
the cut-off function is multiplied furthermore by $\theta(-\epsilon)$.
The strength $G_{\tau}$ is determined for each nucleus such that the
continuous spectrum approximation using the Thomas-Fermi
single-particle level density reproduces the classical empirical
formula $\Delta$ = 12 MeV/${\sqrt{A}}$.

Since the pairing correlation has strong influences on deformations,
one must not trifle the choice of the pairing force
if one wants to perform deformed mean-field calculations.
Our treatment is simple in the sense that it employs the seniority
force but it is an advanced treatment in the sense that the
force strength is determined as a function of the size of the
configuration space for the pairing correlation.

\begin{figure}
\centerline{\includegraphics[angle=-90, width=85mm]{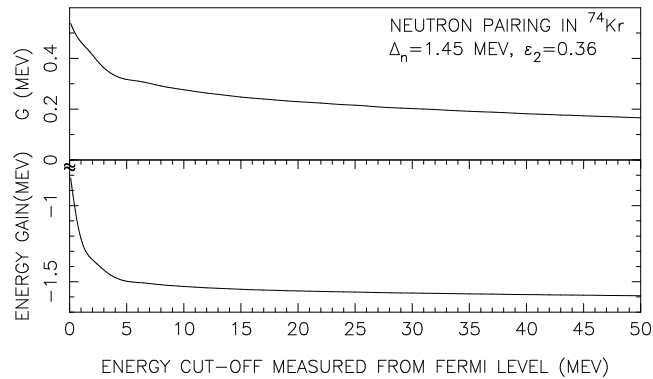}}
\caption{
Dependence of the seniority pairing force strength $G$ renormalized in
such a way that the resulting pairing gap agrees with the experimental
value (top portion) and the dependence of the energy gain due to the
pairing correlation calculated with the strength $G$ (bottom portion) on
the cut-off energy.  Single-particle levels of the Nilsson potential are
used in the BCS calculation.
} \label{g_kr74bcs2}
\end{figure}

It is important not to include positive energy
neutron orbitals in BCS calculations. Otherwise the resulting
density distribution has unphysical neutron gas spreading over
the normalization box.
For neutron-rich nuclei in which the Fermi level is not far from
zero energy, the pairing configuration space
restricted to negative-energy HF orbitals
becomes too small to
obtain a reliable result within the BCS scheme.

Figure~\ref{g_kr74bcs2} shows the results of a test calculation to
illustrate the inadequacy of too small pairing spaces for BCS
calculations.  The calculation is done using the single-particle
spectrum of the Nilsson model, which does not include the continuum
spectrum and free from the problem of the formation of the unphysical
neutron gas.
The strength $G$ of the seniority pairing force is determined in such
a way that, for each value of the cut-off (measured from the Fermi
level), the neutron's pairing gap $\Delta_{\rm n}$ calculated within
the BCS scheme agrees with the experimental value of 1.45 MeV.  The
value of thus defined strength $G$ is shown in the upper half portion
of Fig.~\ref{g_kr74bcs2}.
The bottom half portion shows the energy gain due to
the pairing correlation calculated with the value of $G$ shown in the
top portion. The energy gain does not change sizeably for the cut-off
energy greater than $\sim 4$ MeV. For cut-offs less than that value,
however, the energy gain is non-negligibly smaller than the values
calculated with enough large cut-offs.

\begin{figure}
\centerline{\includegraphics[angle=-90, width=118mm]{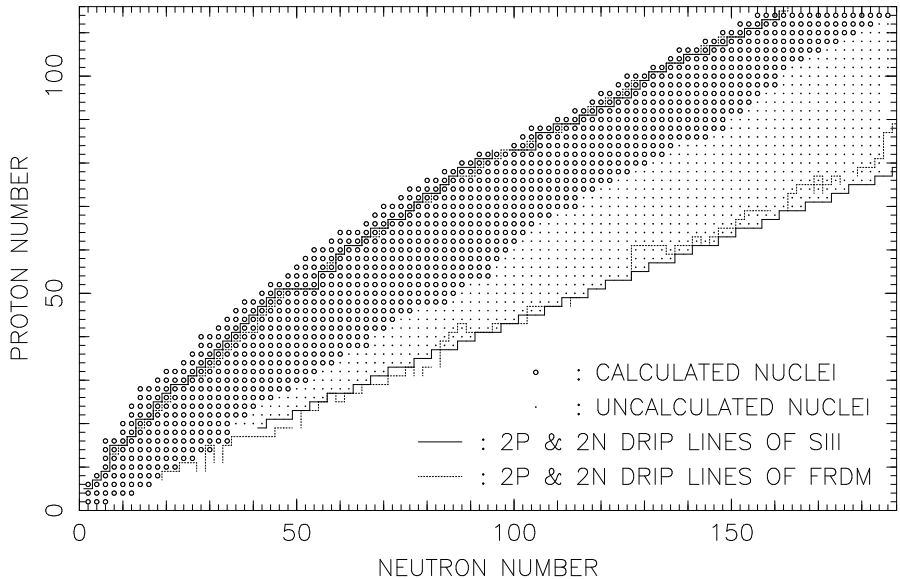}}
\caption{
Even-even nuclei covered by our extensive calculation
(open circles).
Dots are the uncalculated even-even nuclei between the drip lines.
Two-proton and two-neutron drip lines are drawn with solid lines.
Dot lines are those from the results of FRDM\protect\cite{MNM95}.
} \label{g_calnuc}
\end{figure}

From the above consideration concerning the effect
of the continuum states on the
neutron's pairing correlation, it is not very meaningful to apply the
HF+BCS scheme to nuclei in which the Fermi level of neutrons is higher
than $\sim -4$ MeV.  Thus we restrict our calculation within only several
neutrons beyond the experimental neutron-rich frontier.
Figure~\ref{g_calnuc} shows thus selected even-even nuclei.  The
two-proton drip line (the upper one of the solid lines) is deduced
from the total energies of our HF+BCS calculation.  Solution of nuclei
outside the proton drip line can be obtained owing to the Coulomb
barrier.  The two-neutron drip lines (the lower one of the solid
lines) is obtained from the Bethe-Weizs{\"a}cker type mass formula of
Eq.~(\ref{e_BW_mass_formula}) whose four coefficients are determined
through a least-square fitting to the total energies from our HF+BCS
calculation.  One can see that roughly half of the even-even nuclei
are included in our calculation.

In light-mass region, the neutron's Fermi levels of some of the
calculated nuclei are higher than $-4$ MeV.
For such nuclei, the renormalized pairing force strength becomes very
large. We cut the strength at 0.6 MeV.
As a consequence, our solutions for such light-mass nuclei near the
neutron drip line do not have pairing correlation.

The current subject of our research is the calculation of the remaining
half of the nuclear chart on the neutron rich side by developing a
feasible Hartree-Fock-Bogoliubov framework which enables one to
include the effect of the Hartree-Fock continuum (states which are in
the continuum part of the Hartree-Fock single-particle spectrum) on
the pairing correlation.  We think it preferable if the wave functions
are expressed in the coordinate space with a three-dimensional
Cartesian mesh in order to describe possible large surface diffuseness
like thick skins and halos as well as arbitrary deformations.  A
method utilizing the localization of HFB canonical basis seems to be
very promising\cite{Taj98,Taj00a}.


\section{Analyses of the Results}

In this section we reanalyze the results
of our extensive HF+BCS calculations\cite{TTO96},
by presenting the results from new aspects.
Note that all the resulting data of our calculations are available on
the WEB\cite{TTO96}.
In principle one can reproduce the figures given in this section using
only the data on the WEB.

\subsection{Mass}

Concerning the nuclear masses, the r.m.s.\ error of our result turns
out to be 2.2 MeV, which is not so good as those of the latest mass
formulae (typically 0.5 MeV).
Note that this was the first published estimation of the r.m.s.\ mass
error of any of the Skyrme forces.
It is an interesting question how much
the error can be decreased by improving the parameters of the Skyrme
force through an extensive fitting to currently available experimental
masses\cite{TGP00,GTP01}.
To achieve the precision of 100 keV, improvement of
the treatment of the pairing is certainly necessary.
It may also be necessary to add some new terms to the Skyrme force.
Some think that corrections for correlation energies should be included.

Leaving this question which is out of the scope of this paper,
let us reanalyze our result
from different aspects not taken in Ref.~\cite{TTO96}.

\begin{figure}
\centerline{\includegraphics[angle=-90, width=135mm]{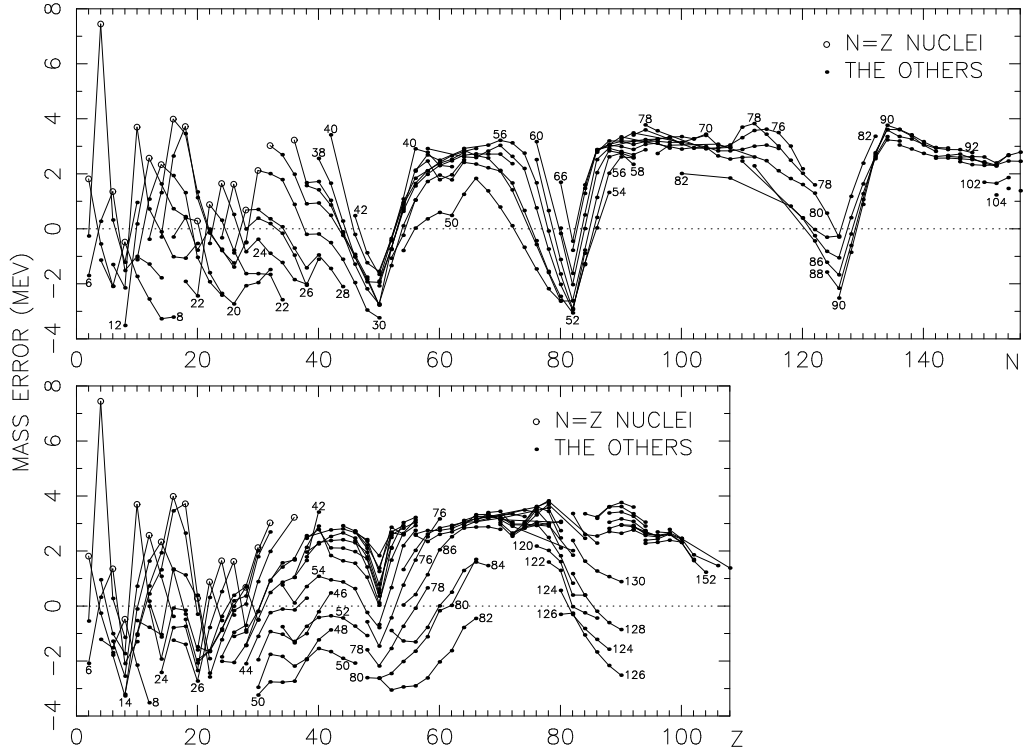}}
\caption{
Errors of the nuclear masses calculated in HF+BCS with Skyrme SIII
force plotted as a function of the neutron number $N$ (top portion)
and the proton number $Z$ (bottom portion). Isotopes and isotones are
connected with lines.  The number printed near each line of isotope or
isotone chain means the proton or neutron number of the chain.
} \label{g_maserr1}
\end{figure}

The top portion of Fig.~\ref{g_maserr1} shows the errors of the
calculated masses as a function of the neutron number $N$.
The bottom portion plots the same data versus the proton number $Z$.
Nuclei of $N=Z$ are designated with open circles
(appearing in $N, Z < 40$) while the others with dots.

One can see that the $N=Z$ nuclei are always at the
highest peaks of the
isotope and the isotone chains to which they belong.  This means that
our mean-field calculation does not have the mechanism corresponding
to the Wigner term of the mass formulae.  It is now widely understood
that the Wigner comes from the $T=0$ pairing. On the other hand, in
the present HF+BCS approach the pairing correlation is considered only
between like nucleons, which means a $T=1$ pairing.

The behavior of the errors looks very systematic in
region $Z, N > 40$ compared with in lighter-mass region.
One can see from the top figure that mass difference becomes
minimum (i.e., the calculated values are more bound than experimental
values) always at neutron's spherical magics of $N$ = 50, 82, and 126.
The situation is less clear in the bottom figure, but at least a
spherical magic of protons, $Z$ = 50, is a minimum of the isotone
chains except for the $N=82$ chain.  This observation suggests that
the employed force has too large a stiffness against deformation.

\begin{figure}
\centerline{\includegraphics[angle=-90, width=77mm]{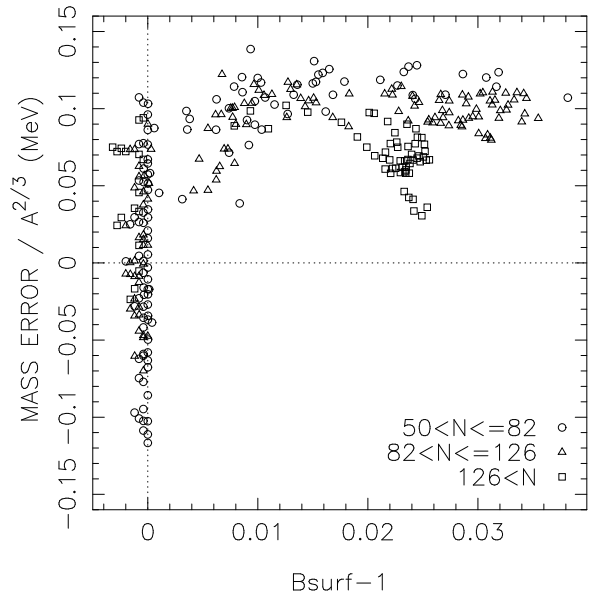}}
\caption{
Errors of nuclear masses divided by a mass-number dependence
factor $A^{2/3}$ as a function of the surface factor $B_{\rm surf}$
subtracted with 1.
Open circles, triangles, and squares are used for the sake of
distinction of the heaviness of the nuclei.
The symbols in the left-hand side of the vertical dot line
$B_{\rm surf}-1=0$ should be regarded as on the line: They are shifted
to the left only to avoid clustering of too many symbols.
} \label{g_maserr2}
\end{figure}

Concerning the possibly too large surface tension of the employed
force, Fig.~\ref{g_maserr2} plots the errors of the calculated nuclear
masses divided by $A^{2/3}$ (which is proportional to the error per
unit surface area) versus $B_{\rm surf} -1$, where $B_{\rm surf}$ is
the ratio of the surface area to that of the spherical shape. The
increase of the surface energy due to deformation is proportional to
$B_{\rm surf} -1$.
The surface area is calculated using liquid-\-drop shape parameters,
whose definition in term of moments is given later in the next subsection.
The error is negative for half of the spherical nuclei ($B_{\rm surf}
= 1$), while for deformed nuclei it is roughly constant and
independent of $B_{\rm surf}$ (i.e., the size of deformation).
The situation is also independent of whether the nucleus is light or
heavy as one can see three kinds of symbols (circles, triangles, and
squares) mixed up in the plot.
These results do not suggest that the tendency that deformed nuclei
have too large masses can be ascribed solely to the surface energy of
the employed force.

\subsection{Deformation}

\begin{figure}
\centerline{\includegraphics[angle=-90, width=75mm]{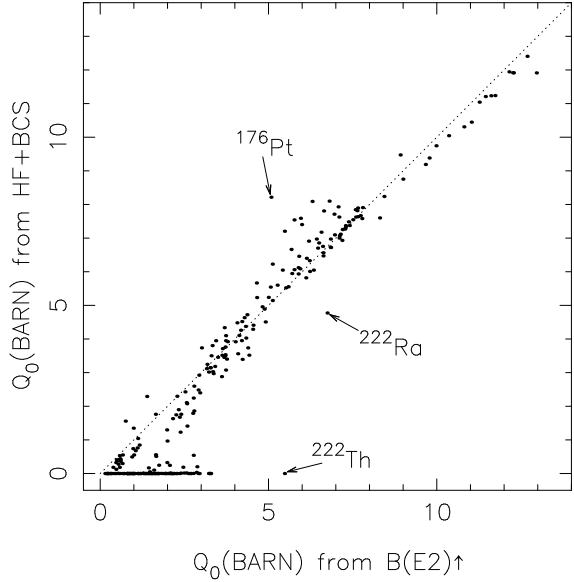}}
\caption{
Comparison of the calculated and experimental quadrupole moment.
} \label{g_qq}
\end{figure}

Fig.~\ref{g_qq} shows a comparison between the experimental and the
calculated intrinsic quadrupole moments.  For each nucleus whose
B(E2; $0^{+} \! \rightarrow \! 2^{+} $) is experimentally known (289
even-even nuclei ranging over 4 $\le Z \le$ 98)\cite{RMM87}, a dot is
marked at a point whose abscissa is equal to the experimental value
and its ordinate to the theoretical value calculated in this paper.
The agreement between the experiment and the theory is excellent in
most cases.  Only a few nuclei exhibit large discrepancies.
Three such nuclei whose element names are indicated in the graph
($^{176}$Pt, $^{222}$Ra, and $^{222}$Th) are around the shape transitional
point of their isotope and isotone chains.  It is relatively difficult
to predict the deformation of shape transitional nuclei.

It has been reported that our results have roughly the same quality as
the other principal theoretical approaches concerning the reproduction
of experimental quadrupole mo\-ments\cite{RSB95}.


Let us define the deformation parameters in terms of multipole moments
for the sake of comparison with the results of other models like FRDM
and also for the easiness to imagine the shapes.
We define the deformation parameters of a HF+BCS solution as those of
a sharp-surface uniform-density liquid drop which has the same mass
moments as the solution has.  The mass density of the liquid drop is
expressed as
\begin{equation} \label{e_radld}
  \rho(\mbf{r}) = \rho_0 \; \theta ( R(\hat{\mbf{r}})-|\mbf{r} |),
  \;\;\;\;\;
  R(\hat{\mbf{r}}) = R_0 \; \Bigl( 1 + \sum_{l=0}^{\infty}
  \sum_{m=-l}^{l}
  a_{lm} Y_{lm} (\hat{\mbf{r}}) \Bigr),
\end{equation}
where $\theta$ is the step (Heaviside) function.
The necessary and sufficient conditions on $a_{lm}$ to fulfill the
reality of $R(\hat{\mbf{r}})$ and the $D_{2h}$ symmetry are that
$l$ and $m$ are even numbers and $a_{lm}=a_{lm}^{\ast}=a_{l-m}$.  We
set $a_{lm}=0$ for $l \ge 6$ and determine the remaining seven
parameters $\rho_0$, $R_0$, $a_{20}$, $a_{22}$, $a_{40}$, $a_{42}$,
and $a_{44}$ such that the liquid drop has the same particle number,
mean-square mass radius, and mass quad\-ru\-pole ($r^2 Y_{2m}$) and
hexa\-deca\-pole ($r^4 Y_{4m}$) moments as the HF+BCS solution has.

By using the mass number $A$ and the mass moments, one obtains the
deformation parameters to express the shape of the mass distribution.
The deformation parameters of proton (neutron) distribution can be
calculated in a similar way by using proton (neutron) number $Z$ ($N$)
and the moments of proton (neutron) distribution.

\begin{figure}
\centerline{\includegraphics[angle=-90, width=135mm]{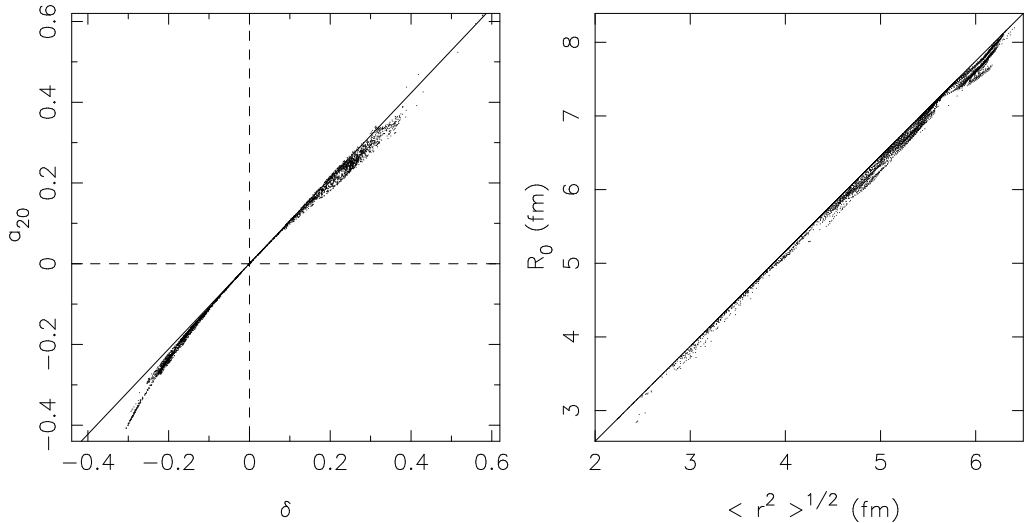}}
\caption{
The correspondence between liquid-drop-model shape parameters and
the related model independent quantities.
The left-hand portion is for the
axially symmetric quadrupole deformation
while the right-hand portion is for the radius.
} \label{g_ldpmom}
\end{figure}

Let us check the adequacy of our definition of the shape parameters by
comparing them with model-independent quantities.

In the left-hand portion of Fig.~\ref{g_ldpmom},
we show the correspondence between two quantities
related to the axially symmetric quadrupole deformation.
They are $a_{20}$ and $\delta$, where the latter is defined as
\begin{equation}
\delta = \frac{3 \langle Q_z \rangle}{4 \langle r^2 \rangle},
\;\;\; Q_z = 2 z^2 - x^2 - y^2,
\;\;\;r^2 = x^2 + y^2 + z^2.
\end{equation}
The number of plotted points is 5361. Each point is for the
deformation of either mass, proton, or neutron distributions of
a HF+BCS solution among 1029 ground and 758 first-excited states.
The solid line represents $a_{20} = (\frac{16}{45}\pi)^{1/2} \delta$,
which is the leading-order expression for sharp-surface density
distributions.
This is a satisfactory result in the sense that
the values of $a_{20}$ are never very distant
from the widely used model independent deformation parameter $\delta$.
We have least-square fitted a polynomial through order three with the
leading order coefficient frozen at the above value. The resulting
polynomial is $a_{20}$ = $(\frac{16}{45}\pi)^{1/2} \delta$ $-0.47
\delta^2$ $+0.78 \delta^3$.  The maximum and the r.m.s.\ deviations
from this polynomial are 0.05 and 0.007, respectively.

In the right-hand portion of Fig.~\ref{g_ldpmom}, we plot
the liquid-\-drop-\-model radius $R_0$
versus the r.m.s.\ radius $r_{\rm rms} = \langle r^2 \rangle^{1/2}$.
The diagonal solid line is $R_0 =  (\frac{5}{3})^{1/2} r_{\rm rms}$,
which is a relation expected to a spherical liquid drop.
It is again a desirable result
that the deviations from this line are never be very large.
Owing to deformations and the surface diffuseness, $R_0$ is likely to
be slightly smaller than $(\frac{5}{3})^{1/2} r_{\rm rms}$.
For the plotted 5361 points, the maximum and the r.m.s.\ deviations
are 0.3 fm and 0.06 fm, respectively.

\begin{figure}
\centerline{\includegraphics[angle=-90, width=135mm]{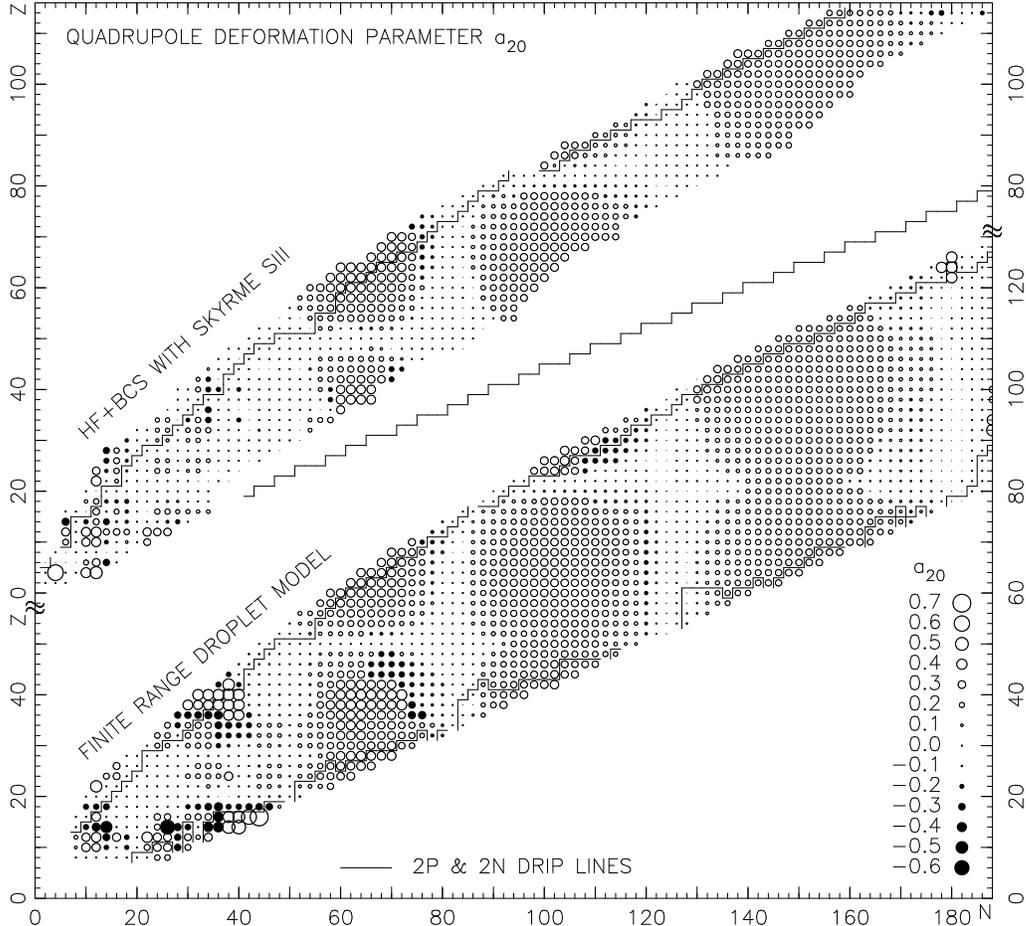}}
\caption{
Comparison of the calculated quadrupole deformation parameter $a_{20}$
between HF+BCS with Skyrme SIII force\protect\cite{TTO96}
and the finite-range droplet model\protect\cite{MNM95}.
The open (solid) circles designate prolate
(oblate) nuclei, while the diameter of the circles is proportional to
the magnitude of the deformation parameter.  The two-proton and
two-neutron drip lines predicted by each model are also drawn.
} \label{g_a20cmp}
\end{figure}

In Fig.~\ref{g_a20cmp}, we compare the axial quadrupole deformation
parameter $a_{20}$ between present HF+BCS calculation and the results
of the FRDM\cite{MNM95}.

A systematic difference which is most easily noticed is found
around $Z \sim N \sim 40$, where the present HF+BCS calculation tends
to predict smaller deformations than the FRDM.
This is due to the shape coexistence prevailing in this region.

\begin{figure}
\centerline{\includegraphics[angle=-90, width=89mm]{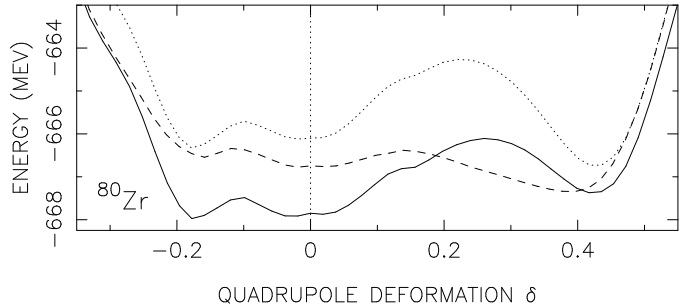}}
\caption{
The Potential energy curve of $^{80}$Zr along the axially symmetric
quadrupole deformation path.
The solid curve is calculated with the SIII force,
the dot curve with the SIII force with 25\% smaller pairing gap,
and the dash curve with the SkM$^{\ast}$ force.
} \label{g_zr80pes}
\end{figure}

An example of the shape coexistence is given in Fig.~\ref{g_zr80pes}.
For $^{80}$Zr, the HF+BCS method with the SIII force (solid curve)
predicts three minima which are energetically competing within $0.6$
MeV.  The order of the energies of these minima can be altered easily
by changing the force parameter to SkM$^{\ast}$\cite{BQB82} (dash
curve) or by decreasing the pairing gap by 25\% (dot curve).

Another systematic difference which one may notice
in Fig.~\ref{g_a20cmp} is in a long and narrow region close
to the proton drip line with $94 \le Z \le 102$, where the FRDM
predicts oblate shapes while our calculations give small-size
prolate shapes.


\begin{figure}
\centerline{\includegraphics[angle=-90, width=87mm]{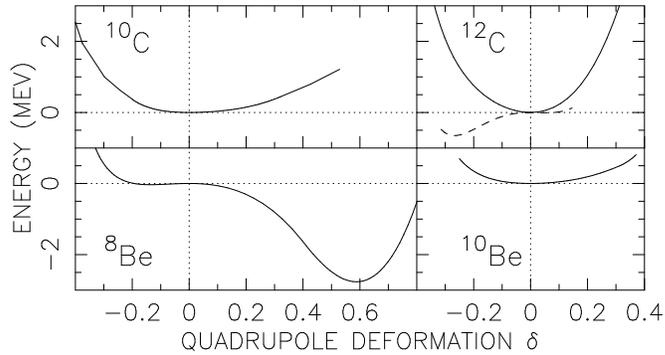}}
\caption{
The potential energy curves for 1p shell even-even nuclei from the
HF+BCS with the SIII force.  The energy is measured from that of the
spherical solution.  For $^{12}$C the results with the Skyrme SII
force is also drawn with dash line.
} \label{g_becpes}
\end{figure}

Figure~\ref{g_becpes} shows the potential energy curves for
$4 \le Z, N \le 6$ nuclei, which are obtained
with a constraint on the axially symmetric mass quadrupole moment
$Q_z$. The abscissa represents the deformation parameter $\delta$.

Among our calculations for the 1029 even-even nuclei,
$^{8}$Be has the largest deformation ($a_{20}$=0.62), which
can be explained in terms of the two-alpha-cluster picture.  In the
bottom-left portion of Fig.~\ref{g_becpes}
one finds the potential energy curve
for this nucleus, which exhibits that the deep prolate minimum is
the only solution (except for the extremely shallow one at an oblate
shape).
%
%

For $^{12}$C, the experimental B(E2)$\uparrow$ is
very large (corresponding to $\vert a_{20} \vert $=0.59) and an oblate
intrinsic deformation with a triangular three-alpha-cluster
configuration has been suggested.  However, the HF+BCS calculation
gives a spherical ground state.  In the top-right portion of the
figure we show the potential energy curve, which has only one minimum
at the spherical shape.  Other widely-used Skyrme forces of the
SkM$^{\ast}$ and the SGII\cite{GS81}
also give the only minimum at sphericity.
On the other hand, calculations using the Nilsson model\cite{Vo74} and
the Strutinski method\cite{LL75} give oblate ground states.
An old Skyrme force SII\cite{Va73} also gives an oblate minimum with
$\delta=-0.27$ (the dash curve in the figure).

Concerning $^{10}$C and $^{10}$Be, the experimental B(E2)$\uparrow$
values indicate large deformations: $\vert a_{20} \vert$=0.82 for
$^{10}$C and 1.1 for $^{10}$Be.  However, the potential energy curves
in Fig.~\ref{g_becpes} have only the spherical minimum.  The quantum
fluctuation in shape may be able to account for the large
B(E2)$\uparrow$ since the curves are very soft toward prolate
deformations.
%
%

It is difficult to conclude definitely only from Fig.~\ref{g_becpes}
what deformation $^{12}$C should bear in the HF approximation, since
the optimal shapes of light nuclei are apt to be changed when effects
beyond mean-field approximations are taken into account.  The parity
projection has been reported to be especially important\cite{KH95}
because the triangular three-alpha-cluster configuration violates the
symmetry.

Let us briefly report on the three-alpha-cluster linear-chain state of
$^{12}$C. By extending the potential energy curve in
Fig.~\ref{g_becpes} to larger $\delta$, we have found the first
excited minimum at $\delta \sim 1.0$, in good agreement with the
result of the Nilsson model for the linear-chain state
($\delta$=1.1)\cite{ZZC91}.  The excitation energy from our
calculation is 21 MeV. Though it is much larger than the experimental
value of 7.654 MeV, the overestimation will be improved by the angular
momentum projection.
%
%

\begin{figure}
\centerline{\includegraphics[angle=-90, width=135mm]{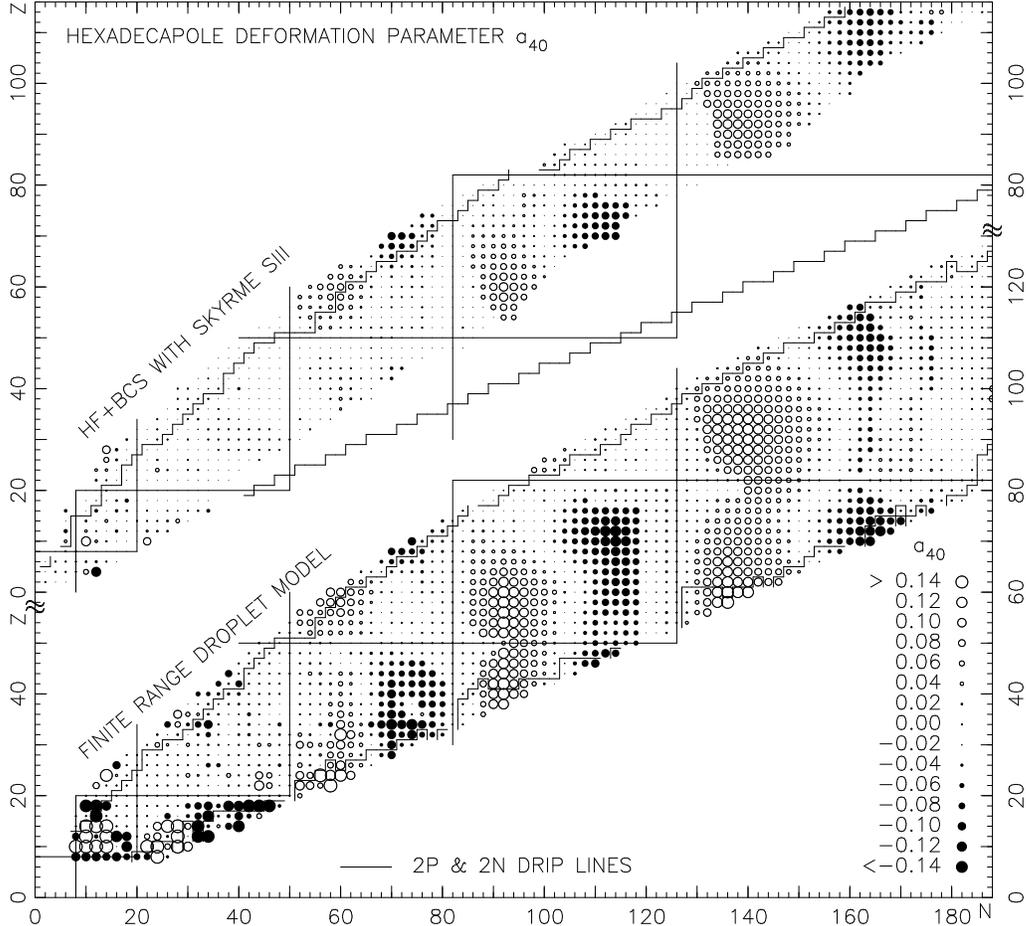}}
\caption{
%
%
Same as in Fig.~\protect\ref{g_a20cmp}, but for the hexadecapole
deformation parameter $a_{40}$.  The grid indicates the locations of
the magic numbers for spherical shapes.
The results of FRDM is taken from Ref.~\protect\cite{MNM95}.
} \label{g_a40cmp}
\end{figure}

In Fig.~\ref{g_a40cmp}, we compare the axially symmetric hexadecapole
deformation parameter $a_{40}$ between the present HF+BCS calculation
and the FRDM result\cite{MNM95}.
A common feature between the results of the two models is that the
sign of $a_{40}$ is positive in the bottom-left quarter of the major
shells and negative in the top-right quarter.
This behavior is perfectly understandable in terms of the density
distribution of pure-$j$ single-particle wave functions.
A systematic difference between two calculations is found in $Z , N <
50$ where $|a_{40}|$ is very small with exception of only several nuclei
in the HF+BCS while it has certain sizes in many nuclei according to the
FRDM.
In particular, most of the light nuclei with $N, Z < 20$ have
conspicuously large $|a_{40}|$ in FRDM.
For heavier nuclei, too, the magnitude of $a_{40}$ is generally smaller
in HF+BCS than in FRDM.
Further investigation is necessary to see whether these
differences are originated in the different definitions of the shape
parameters: In the HF+BCS they are calculated from the moments, while
in the FRDM they are the input parameters to specify the shape of the
single-particle potential.


One of the advantages of mean-field methods over shell-correction
schemes is that the protons and the neutrons do not have to possess
the same radius and deformation.
Indeed, it is one of the reasons to adopt the Cartesian-mesh
representation rather than the expansion in harmonic-oscillator bases.
In order to make the best use of this advantage, we have calculated
the liquid-\-drop shape parameters separately for protons and neutrons
for 1029 ground and 758 first-excited solutions.

\begin{figure}
\centerline{\includegraphics[angle=-90, width=124mm]{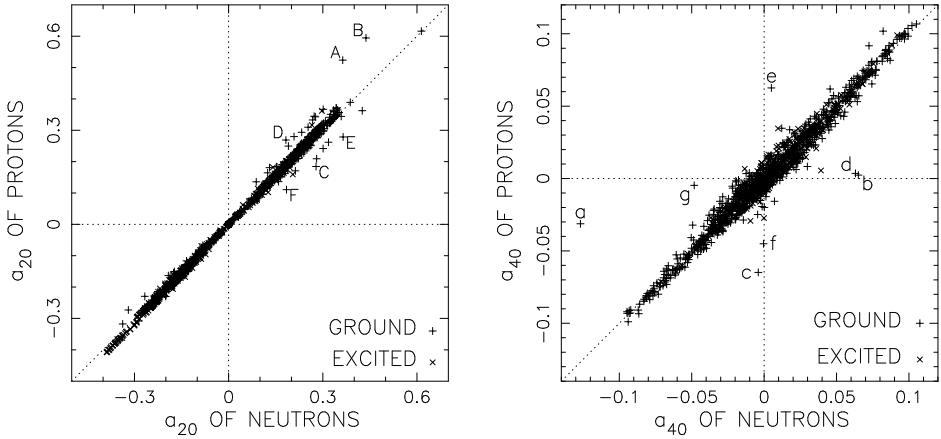}}
\caption{
Difference of the shape parameters between protons and neutrons for
the ground states of 1029 even-even nuclei (denoted by plus symbols)
and the first-excited solutions (denoted by cross symbols) of 758
even-even nuclei for HF+BCS with SIII force.  The left and the right
portions show respectively the quadrupole deformation parameter
$a_{20}$ and the hexa\-deca\-pole deformation parameter $a_{40}$.
} \label{g_almpn}
\end{figure}

The left-hand portion of Fig.~\ref{g_almpn} presents a comparison of
the values of $a_{20}$ between protons and neutrons.
The r.m.s.\ value of the difference
$|a_{20}^{\rm neu} - a_{20}^{\rm pro}|$
is 0.012, while the maximum difference occurs
in the ground state of $^{14}_{\phantom{0}4}$Be$_{10}$
($a_{20}^{\rm neu}=0.36$, $a_{20}^{\rm pro}=0.52$), which is indicated
by letter {\bf A} in the figure.  Symbols indicated by letters from
{\bf B} to {\bf F} are the ground states of
$^{16}_{\phantom{0}4}$Be$_{12}$
($a_{20}^{\rm neu}=0.44$, $a_{20}^{\rm pro}=0.59$),
$^{30}_{18}$Ar$_{12}$
($                 0.28$, $                 0.18$),
$^{30}_{12}$Mg$_{18}$
($                 0.18$, $                 0.27$),
$^{36}_{24}$Cr$_{12}$
($                 0.36$, $                 0.28$), and
$^{28}_{18}$Ar$_{10}$
($                 0.18$, $                 0.11$), respectively.
Among these nuclei, $^{30}$Ar, $^{36}$Cr, and $^{28}$Ar are outside
the proton drip line.  The absolute value of the difference is smaller
than 0.05 for 98.9 \% of the solutions.

The right-hand portion of Fig.~\ref{g_almpn} is similar to the
left-hand portion but for $a_{40}$.
The r.m.s.\ value of the difference
$|a_{40}^{\rm neu} - a_{40}^{\rm pro}|$ is 0.0063,
while the maximum difference is found in $^{16}_{\phantom{0}4}$Be$_{12}$
($a_{40}^{\rm neu}=-0.127$, $a_{40}^{\rm pro}=-0.031$),
which is indicated with letter {\bf a}.
Symbols indicated by letters from {\bf b}
to {\bf g} are the ground state solutions of
$^{14}_{\phantom{0}4}$Be$_{10}$
($a_{40}^{\rm neu}=0.065$, $a_{40}^{\rm pro}=0.002$),
$^{18}_{12}$Mg$_{6}$
($                -0.004$, $                -0.065$),
$^{22}_{12}$Mg$_{10}$
($                 0.063$, $                 0.004$),
$^{22}_{10}$Ne$_{12}$
($                 0.005$, $                 0.063$),
$^{22}_{16}$S$_{6}$
($                 0.000$, $                -0.045$), and
$^{18}_{\phantom{0}6}$C$_{12}$
($                -0.048$, $                -0.005$), respectively.
Among these nuclei, $^{18}$Mg and $^{22}$S are outside the proton drip
line.

To summarize,
according to the results of our calculations with the Skyrme SIII force,
the largest differences between the shapes of proton and neutron density
distributions  occur in the lightest nuclei.
Except for these nuclei, the differences are not remarkably large.

\begin{figure}
\centerline{\includegraphics[angle=-90, width=125mm]{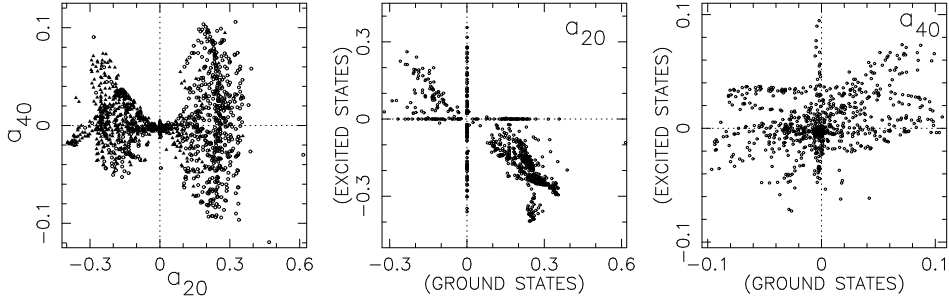}}
\caption{
Correlation between quadrupole and hexadecapole deformation parameters.
The left-hand portion shows the relation between $a_{20}$ and $a_{40}$
of each HF+BCS solution. Open circles are for ground state solutions
while solid triangles are for the first excited solutions.
The middle portion shows the relation between
$a_{20}$ of the ground state and $a_{20}$ of the first excited state
of a nucleus.
An open circle corresponds to a nucleus.
Only a kind of symbol is used.
Clustering open circles may look like solid circles.
The right portion is the same as the middle portion
except it is for $a_{40}$.
} \label{g_almcor}
\end{figure}

In Fig.~\ref{g_almcor}, we investigate the correlations between
deformation parameters.
The left-hand portion of the figure shows that the
ground states (open circles) are mostly with $0.2 < a_{20} < 0.3$
while the excited states (solid triangles) are with
$-0.3 < a_{20} < 0$.
For prolate ($a_{20}>0$) states $a_{40}$ ranges from $-0.1$ to $0.1$
while for oblate ($a_{20} <0$) states
$a_{40}$ covers a smaller interval from $-0.05$ to $0.07$.
For nuclei with $a_{20} \sim 0$, $a_{40}$ is also close to zero.

The middle portion of Fig.~\ref{g_almcor} shows that $a_{20}$ of the
ground state of a nucleus and $a_{20}$ of the first excited state of
the same nucleus have a strong negative correlation. The most frequent
case is that the ground state is prolate and the first excited state
is oblate.  There are opposite cases, too, but the number is much less.
There are also many nuclei in which either ground or the first excited
states is spherical ($a_{20}=0$).

The right-hand portion of Fig.~\ref{g_almcor} shows that $a_{40}$ have
more complicated correlation pattern between ground and the first
excited states of a nucleus.

\begin{figure}
\centerline{\includegraphics[angle=0, width=125mm]{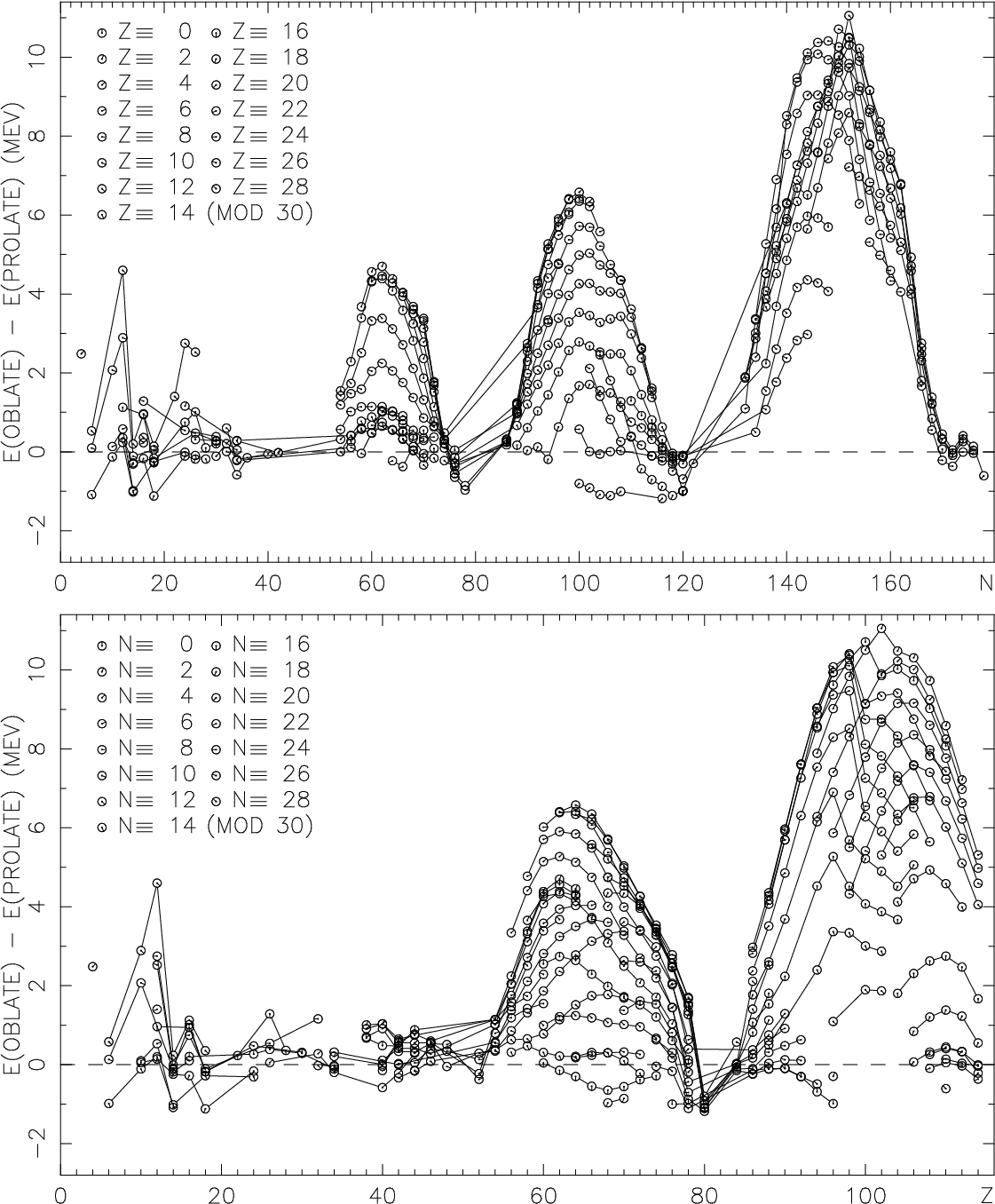}}
\caption{
Energy difference between prolate and oblate solutions.
Top and bottom portions show the same results for different abscissae.
In the top portion, the abscissa is the neutron number
and isotopes are connected with lines.
It is possible to read off the proton number of each isotope chain
from the direction of the ``hand'' put in each circle.
See the legends in the top-left corner of each portion.
In the bottom portion, the abscissa is the proton number
and isotones are connected with lines.
} \label{g_oblpro2}
\end{figure}

A merit of extensive calculations is that it allows one to see global
trends of nuclear properties over the entire area of the nuclear
chart.
As another example of the analysis of such a global trend, we have
investigated the systematics of the energy difference between the
oblate and the prolate solutions.
In the top portion of Fig.~\ref{g_oblpro2},
the energy difference is plotted versus the neutron number.
One can see an evident difference between below and above
the $N=50$ shell magic: For nuclei with $N < 50$, the oblate solutions
often have lower energies than prolate ones. For nuclei with $N > 50$,
oblate ground states are very rare and found only in nuclei
whose $N$ is slightly smaller than a magic number.
In the bottom portion, the same data is plotted versus the proton
number $Z$.  One can see that the situation is exactly the same for
proton's shell effect.

This abrupt change at $N=50$ and $Z=50$ seems to suggest that the
dominance of prolate deformations in rare-earth and actinide nuclei
may be related with the nature of the Mayer-Jensen major shell while
the harmonic-oscillator major shell leads to an equal situation
between oblate and prolate deformations.
The Mayer-Jensen major shell is composed of the normal parity orbitals
which have the same oscillator quantum number and a unique parity
orbital which is pushed down from the next oscillator shell by the
spin-orbit potential due to its large angular momentum.
We have suggested\cite{TTO96} that one may ascribe the prolate
dominance to the spin-orbit potential.

Recently, we have performed Nilsson-Strutinsky calculations to answer
this question\cite{TS01}.  The Nilsson potential contains a $l^2$
potential, which simulates a square-well like radial profile of the
mean potential, and a spin-orbit potential.  The $l^2$ potential can
reproduce the prolate dominance without the help of the spin-orbit
potential, while the spin-orbit potential alone cannot reproduce the
prolate dominance.  However, this does not mean the weakness of the
effect of the spin-orbit potential.  On the contrary, it interferes so
strongly with the $l^2$ potential that the ratio of the number of
prolate nuclei to that of oblate ones oscillates with a large
amplitude, making a prolate dominance with the standard strength,
reverting the situation to a oblate-favor one with half the standard
strength, and again giving rise to a prolate dominance with vanishing
strength.

\subsection{Radius}

\begin{figure}
\centerline{\includegraphics[angle=-90, width=115mm]{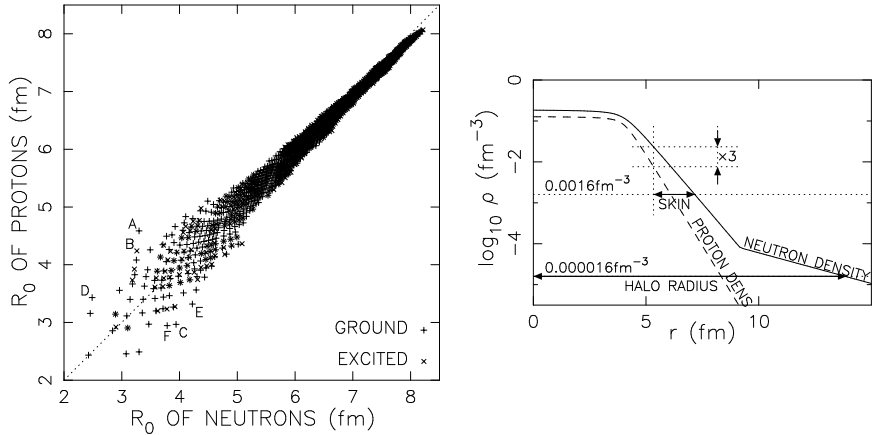}}
\caption{
The left-hand portion shows a comparison of the liquid-\-drop radius
$R_0$ of the proton distribution and that of the neutron distribution.
The right-hand portion shows the definitions employed in this paper of
the skin thickness and the halo radius.
}
\label{g_radpn}
\end{figure}

The left-hand portion of Fig.~\ref{g_radpn} presents a comparison of
the liquid-\-drop radius of protons $R_{0}^{\rm pro}$ with that of
neutrons $R_{0}^{\rm neu}$.

\noindent
The r.m.s.\ value of the difference
$|R_{0}^{\rm neu}-R_{0}^{\rm pro}|$ is 0.18 fm.
Symbols indicated by letters from {\bf A} to {\bf F} are the solutions of
$^{22}_{16}$S$_{6}$
($R_0^{\rm neu}=3.3$ fm, $R_0^{\rm pro}=4.6$ fm),
$^{20}_{14}$Si$_{6}$
($              3.3$, $              4.2$),
$^{16}_{\phantom{0}4}$Be$_{12}$
($              3.9$, $              3.0$),
$^{8}_{6}$C$_{2}$
($              2.5$, $              3.4$),
$^{22}_{\phantom{0}6}$C$_{16}$
($              4.2$, $              3.3$), and
$^{14}_{\phantom{0}4}$Be$_{10}$
($              3.8$, $              2.9$), respectively.
Among the nuclei mentioned above, $^{20}$Si is for the excited state
while the others are for the ground states.  $^{22}$S, $^{20}$Si, and
$^8$C are outside the proton drip line.  Hence the maximum difference
for nuclei inside the drip lines occurs in the ground state of
$^{16}$Be.
Incidentally, when we plotted the r.m.s.\ radii $r_{\rm rms}$
instead of $R_0$, we found that the resulting graph looked very
similar to this figure except for a constant factor, $R_0$ $\sim$
$(\frac{5}{3})^{1/2}$ $r_{\rm rms}$.

Concerning the difference of radial density distribution between
protons and neutrons, one should pay attention not only to volume
properties represented by liquid-\-drop or r.m.s.\ radius but also to
surface properties, which are sensitive to the difference at much
lower densities and manifest themselves, e.g.,
in the thicknesses of nucleon skins and the radii of nucleon halos.
The right-hand portion of Fig.~\ref{g_radpn} is an illustration
to explain our definitions of these two quantities.
We regard that
a point is in the proton skin if
$\rho_{\rm p} > 3 \rho_{\rm n}$ and
$\rho_{\rm p} + \rho_{\rm n} > 0.0016 \mbox{fm}^{-3}$,
where $\rho_{\rm p}$ and $\rho_{\rm n}$ are respectively
the proton density and the neutron density at that point.
A similar definition has been proposed in Ref.~\cite{FOT93}
The halo radius is defined as the radius at which
angle-averaged mass density is $1.6 \times 10^{-5}$ fm$^{-3}$.

\begin{figure}
\centerline{\includegraphics[angle=-90, width=123mm]{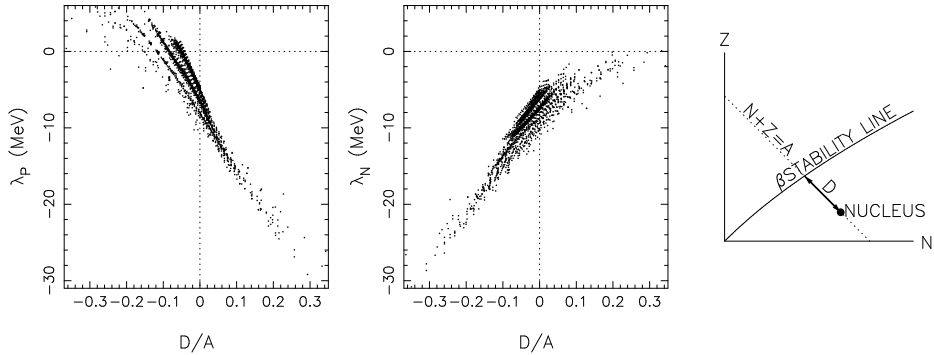}}
\caption{
Fermi levels as functions of the distance from the $\beta$-stability
line $D$ divided by the mass number $A$.
The left and the middle portions are for protons and neutrons,
respectively.
The right portion explains the definition of $D$.
} \label{g_fermi}
\end{figure}

As a preparation before discussing about skins and halos,
let us first look at the relation between the Fermi levels and
the distance in the ($N,Z$) plane from the $\beta$-stability line.
As illustrated in the right-hand portion of Fig.~\ref{g_fermi},
we choose to measure the distance $D$
along a line of constant mass number.
We assign a negative sign to $D$ for nuclei with $Z > N$.
As the definition of the $\beta$-stability line,
we employ a classical empirical formula,
\begin{equation}
  N - Z = \frac{0.4 A^2}{A+200}, \;\;\; A=N+Z.
\end{equation}
The left- and right-hand portions of Fig.~\ref{g_fermi} show
respectively the Fermi level of the protons $\lambda_{\rm p}$
and that of the neutrons $\lambda_{\rm n}$.
Both ground-state and first excited solutions are plotted.
The abscissa is the distance $D$ divided by the mass number $A$.
One can see a strong correlation in both portions.
(Incidentally, the dots spread over larger area if the
the abscissa is changed to $DA^{-2/3}$ or
$DA^{-4/3}$.)
These figures suggest that one can
measure the unstableness of a nucleus in
two roughly equivalent ways.
One may ask how high one of the Fermi levels is instead of asking how
far the nucleus is located from the $\beta$-stability line in the
nuclear chart.

Now, by using the next two figures, we will show that
the skin grows mo\-not\-o\-nous\-ly and regularly as nucleons are added to the
nucleus while the halo radius grows very slowly except near the drip
lines, where it changes the behavior completely and expands very
rapidly.

\begin{figure}
\centerline{\includegraphics[angle=-90, width=114mm]{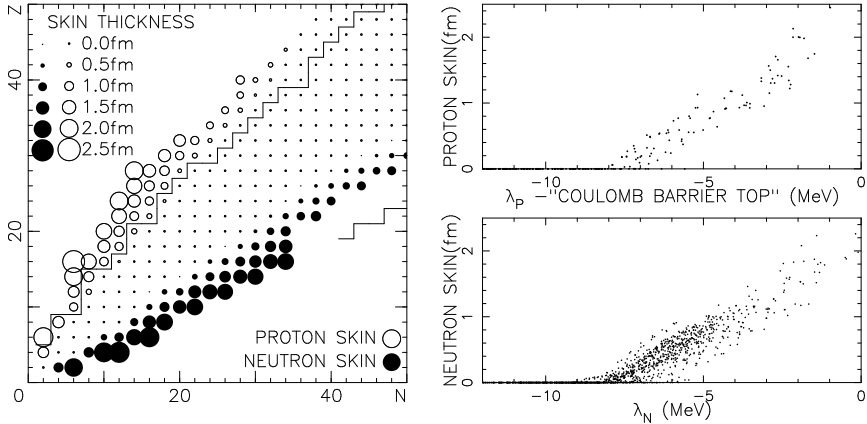}}
\caption{
Skin thickness.  In the left portion the thickness of the proton
(neutron) skin of the ground states
is shown with the radius of open (solid) circles.
Nuclei without either type of the skins are shown with the smallest
dots.
The staircase like lines represent two-proton and two-neutron drip
lines.
The right top
portion shows the thickness of proton skin as a function of
proton's Fermi level measured from the top of the Coulomb barrier
as defined in the text.
The right bottom portion shows the thickness of neutron skin versus
neutron's Fermi level.
} \label{g_skin2}
\end{figure}

The left-hand portion of Fig.~\ref{g_skin2} shows
the thickness of the proton and neutron skins in the ($N, Z$) plane.
%
%
Note that the absolute magnitudes
of this quantity depend on our specific
choice of the definition of the skins but the relative differences
between adjacent nuclei are rather insensitive to the definition.
The increase of skin thickness along isotope and isotone chains
looks monotonous.
If one would fold the graph in a line $N=Z$, one could easily find a
general rule that the proton skin of a proton rich nucleus is larger
than the neutron skin of its mirror nucleus.
This fact can be ascribed to the Coulomb repulsion between protons.
However, since the proton drip line is closer to the $N=Z$ line than
the neutron drip line is, proton skins are much rare than neutron
skins if nuclei outside the drip lines are excluded.

The top-right and the bottom-right portions of Fig.~\ref{g_skin2}
present the skin thickness of protons and neutrons, respectively.
The Fermi levels are used as the abscissae.  These figures include the
data of not only those nuclei shown in the figure in the left-hand
portion but all the calculated 1029 even-even nuclei.
They also include not only the ground-state solutions but also
first-excited solutions for 758 nuclei.
Most of the dots are on the bottom line.
From the figure in the right-bottom portion, one can see a general
rule that the neutron's skin (by our definition) appears for
  $\lambda_{\rm n} > -8 \mbox{MeV}$
and increases linearly with $\lambda_{\rm n}$ at a rate of
$\sim 0.3$fm/MeV.
This rule is rather insensitive to the mass and
applies both to ground and excited states.

For protons, the Fermi level should be measured from the top of the
Coulomb barrier.  Otherwise the dots do not cluster in a narrow
area.
%
%
We estimate the height of the barrier top as
\begin{equation}
E_{\rm C.B.} = \frac{1.44 Z}{1.2 A^{1/3} + \Delta R} \; \mbox{(MeV)},
\;\;\; \Delta R = 1.0\;\mbox{fm},
\end{equation}
where $\Delta R$ is a parameter to designate the shift of the location
of the barrier top from the liquid-\-drop surface.
We assume a constant value of $\Delta R$ for all the nuclei.
Small changes of $\Delta R$ only translate uniformly the plotted dots.
We choose $\Delta R = 1.0$ fm so that the horizontal location
where the dots depart form the bottom line is roughly the same
as in the graph for neutrons.
The resulting behavior of the skin thickness of protons
is very similar to that of the neutrons.

\begin{figure}
\centerline{\includegraphics[angle=-90, width=103mm]{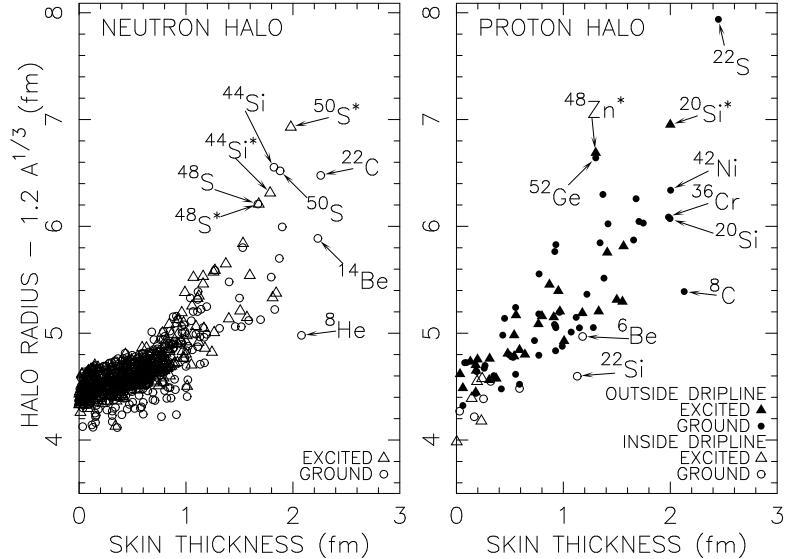}}
\caption{
The radius of the neutron (proton) halo plotted versus the thickness of the
neutron (proton) skin (in the left (right) portion).
Element names are printed for nuclei with large halo radius.
An asterisk symbol put to the right shoulder of the element name indicates
that the state is the first excited solution of the nucleus.
} \label{g_halo}
\end{figure}

Fig.~\ref{g_halo} shows ``the thickness of the halo'', i.e., the radius
of the halo subtracted with the liquid-\-drop radius $1.2 A^{-1/3}$ fm.
The abscissa is the thickness of neutron (proton) skin for the neutron
(proton) halo.
One can see that halos grow only slowly for skin thickness less than 1
fm but it expands rapidly after 1 fm.
This sudden change is widely understood to be ascribed to the last
occupied orbital's spatial extension due to its small binding energy.

%
%
%


\section*{Acknowledgements}

The author thanks Dr.~P.~Bonche, Dr.~H.~Flocard, and Dr.~P.-H.~Heenen
for providing a Cartesian-mesh HF+BCS code {\sl EV8}\cite{BFH85}.
They are also grateful to Dr.~T.~Tachibana and Dr.~K.~Oyamatsu for the
TUYY mass formula code, and to Dr.~S.~Raman for the computer file of
the B(E2)$\uparrow$ table.
Most of the calculations discussed in this paper have been carried out
in collaboration with Dr.~S.~Takahara of Kyorin University.

\appendix

\section{Cartesian Mesh and the Fourier Basis} \label{a_fourierbasis}

In this Appendix we explain the origin of the surprising high accuracy
of nuclear mean-field calculations with apparently coarse Cartesian
meshes.
It is sufficient to consider a one-dimensional space
because the essence is there.
We put $N$ mesh points at $x_i = (i - \frac{1}{2})a$ ($1 \le i \le N$)
in an interval $0 \le x \le N a$ as in the following illustration.

%
%
\setlength{\unitlength}{2cm}
\noindent
\begin{picture}(6.5,1)(-0.5,-0.5)
\multiput(0,-1)(1,0){6}{
\begin{picture}(1,2)(0,-1)
\put(0,0){\line(1,0){1}}
\put(0,-0.05){\line(0,1){0.1}}
\put(1,-0.05){\line(0,1){0.1}}
\put(0.5,0){\circle*{0.05}}
\put(0.0283,0.0225){\circle*0.01} \put(0.0576,0.0437){\circle*0.01}
\put(0.0879,0.0635){\circle*0.01} \put(0.1191,0.0818){\circle*0.01}
\put(0.1511,0.0987){\circle*0.01} \put(0.1838,0.1141){\circle*0.01}
\put(0.2172,0.1280){\circle*0.01} \put(0.2512,0.1403){\circle*0.01}
\put(0.2857,0.1511){\circle*0.01} \put(0.3207,0.1602){\circle*0.01}
\put(0.3561,0.1677){\circle*0.01} \put(0.3918,0.1735){\circle*0.01}
\put(0.4278,0.1777){\circle*0.01} \put(0.4638,0.1802){\circle*0.01}
\put(0.5000,0.1810){\circle*0.01} \put(0.5362,0.1802){\circle*0.01}
\put(0.5722,0.1777){\circle*0.01} \put(0.6082,0.1735){\circle*0.01}
\put(0.6439,0.1677){\circle*0.01} \put(0.6793,0.1602){\circle*0.01}
\put(0.7143,0.1511){\circle*0.01} \put(0.7488,0.1403){\circle*0.01}
\put(0.7828,0.1280){\circle*0.01} \put(0.8162,0.1141){\circle*0.01}
\put(0.8489,0.0987){\circle*0.01} \put(0.8809,0.0818){\circle*0.01}
\put(0.9121,0.0635){\circle*0.01} \put(0.9424,0.0437){\circle*0.01}
\put(0.9717,0.0225){\circle*0.01}
\end{picture}
}
\put(0.15,-0.4){\makebox(0.9,0.3)[t]{$x_1$}}
\put(1.15,-0.4){\makebox(0.9,0.3)[t]{$x_2$}}
\put(2.15,-0.4){\makebox(0.9,0.3)[t]{$x_3$}}
\put(3.15,-0.4){\makebox(0.9,0.3)[t]{$x_4$}}
\put(4.15,-0.4){\makebox(0.9,0.3)[t]{$\cdots$}}
\put(5.15,-0.4){\makebox(0.9,0.3)[t]{$x_N$}}
\multiput(0.15,0.25)(1,0){4}{\makebox(0.9,0.3)[b]{$a$}}
\put(4.15,0.25){\makebox(0.9,0.3)[b]{$\cdots$}}
\put(5.15,0.25){\makebox(0.9,0.3)[b]{$a$}}
\end{picture}
%

\noindent
In the mesh representation, any function $\psi(x)$ defined in this
interval is expressed in terms of a set of its values at the mesh
points $\psi_i = \psi(x_i)$.
The equation to determine $\{ \psi_i \}$
is usually derived as a discrete approximation to the Schr{\"o}dinger
equation.
An alternative point of view was presented by Baye and Heenen\cite{BH86}.
They introduced a set of
orthogonal basis functions $f_i(x)$,
\begin{equation} \label{eq_fi}
f_i(x) = f \left( \frac{x-x_i}{a} \right), \;\;\;\;\;
f(\xi)=\frac{\sin N a k_0 \xi}{N \sin a k_0 \xi},
\;\;\;\;\; k_0=\frac{\pi}{N a},
\;\;\;\;\; 1 \le i \le N.
\end{equation}
%
%
\begin{figure}
\centerline{\includegraphics[angle=-90, width=84mm]{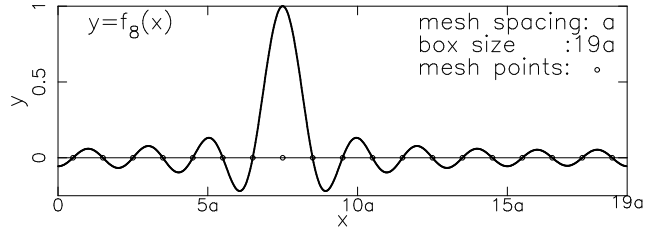}}
\caption{
The graph of a Fourier basis function $y=f_8(x)$ corresponding to the
8th mesh point for mesh spacing $a$ and interval $19a$.  All the mesh
points are indicated with open circles on the $y=0$ line.
} \label{g_fbfunc}
\end{figure}
%
The shape of the above basis function is shown
in Fig.~\ref{g_fbfunc}.
When $\psi(x)$ can be expressed as a linear combination of $\{ f_i(x) \}$,
$\psi_i$ is nothing but the coefficient of $f_i(x)$ ,
\begin{equation} \label{eq_twf}
\psi(x)=\sum_{i=1}^{N} \psi_i f_i(x),
\end{equation}
owing to a property of the basis, $f_i(x_j)=\delta_{ij}$.
From this point of view, one can uniquely derive the equation to
determine $\{ \psi_i \}$ from the variational principle in a
space spanned by the trial wave functions of form (\ref{eq_twf})
and  can avoid
the arbitrariness to choose a discrete approximation formula.

The expression (\ref{eq_twf}) is equivalent
to a truncated Fourier expansion with a boundary condition
$\psi(x+Na)$ = $(-1)^{N+1}\psi(x)$ because one can write
\begin{equation}
f_j(x)=\frac{1}{N} \sum_{n=1}^{N} e^{i (2n-N-1) k_0 (x-x_j)}.
\end{equation}
This is the very reason why the Cartesian mesh representation provides
high accuracy in nuclear calculations: The Cartesian mesh is
associated with the Fourier or plane-wave basis, which is especially
suitable to the atomic nucleus owing to its saturation property.

Concerning integrals, the mid-point rule is exact for the product of
two wave functions,
\begin{equation} \label{e_mid_point_rule}
\int_0^{Na} \psi^{\ast} (x) \phi(x) d x =
a \sum_{i} \psi^{\ast}_{i} \phi_{i},
\end{equation}
where $\psi(x)$ and $\phi(x)$ are functions of form (\ref{eq_twf})
and $\phi_i = \phi(x_i)$.
For other integrands like $\psi^{\ast} (x) V(x) \phi(x)$,
the mid-point rule similar to Eq.~(\ref{e_mid_point_rule}) is not
exact any more.
Nevertheless, one should yet use the mid-point formula rather than
high-order formulae like Eq.~(\ref{e_eNC_integ})
for the sake of high precision.

On the other hand, the expressions for derivatives are not so simple.
The expression of the $\nu$th derivative of
the function defined by Eq.~(\ref{eq_twf}) at a mesh point $x_i$
involves a full $N \times N$ matrix $D^{(\nu)}_{ij}$:
\begin{equation} \label{eq_deriv}
\left. \frac{d^{\nu}}{dx^{\nu}} \psi(x) \right\vert_{x=x_i}   =
\frac{1}{a^{\nu}} \sum_{j=1}^{N} D^{(\nu)}_{ij} \psi_{j},
\end{equation}
where the matrices for $\nu$ = 1 and 2 are expressed respectively as
\begin{equation} \label{e_first_differential_matrix}
D^{(1)}_{ij} =
  \left\{
    \begin{array}{cc}
      0 & (i=j)\\
      {\displaystyle (-1)^{i-j}\frac{\pi}{N \sin \frac{i-j}{N}\pi}}
      & (i \not= j)\\
    \end{array}
  \right.
\end{equation}
and
\begin{equation} \label{e_second_differential_matrix}
D^{(2)}_{ij} =
  \left\{
    \begin{array}{cc}
      {\displaystyle \frac{1-N^2}{3N^2}\pi^2 } & (i=j)\\
      {\displaystyle -(-1)^{i-j}
      \frac{2\pi^2}{N^2 \tan \frac{i-j}{N}\pi \sin \frac{i-j}{N}\pi}}
      & (i \not= j)
    \end{array}
  \right. \;\;\; .
\end{equation}
Since the evaluation of this expression seems to require a rather long
computation time for the resulting precision,
the code {\sl EV8}\cite{BFH85} employs a finite-point
approximation formula like Eq.~(\ref{eq:numderiv}) to
approximate Eq.~(\ref{eq_deriv}) rather than sticking to the
variational picture as completely as possible.

The extension to the three-dimensional case is straightforward. A
wave function is expressed by its values $\psi_{ijk}$ at mesh points
$(x_i,y_j,z_k)$ =
$((i-\frac{1}{2})a,(j-\frac{1}{2})a,(k-\frac{1}{2})a)$ as,
\begin{equation} \label{eq_Fou3d}
\psi(x,y,z)= \sum_{ijk} \psi_{ijk} f_i(x) f_j(y) f_k(z).
\end{equation}

It is worth noting that, for polar and cylindrical coordinates, the
mesh points become non-uniform in order to make the associated bases
the eigenfunctions of the kinetic energy.  If one takes the
radial mesh points equidistantly, the spacing must be finer
than that of the three-dimensional Cartesian mesh.
Meshes with 1 fm spacing are too coarse for the
radius grid but fine enough for three-dimensional Cartesian mesh.


\section{Finite-Point Formulae for Numerical Differentiation}

The $\nu$th derivative of a function $f(x)$ can be evaluated
approximately by the values of the function at $n$ mesh points as,
\begin{equation} \label{eq:numderiv}
\frac{d^{\nu}f(x)}{dx^{\nu}} = \frac{1}{b_n^{(\nu)} a^{\nu}}
\sum_{i=(n-1)/2}^{(n-1)/2} c_{ni}^{(\nu)} f(x+ai)
+ {\cal O}(a^{n-[\nu-1]_{\rm e}-1}),
\end{equation}
where $n \ge 3$ is an odd number while
$[j]_{\rm e}$ means the maximum even number not exceeding $j$.
The coefficients $c_{ni}^{(\nu)}$ and $b_n^{(\nu)}$
can be determined by approximating
$f(x)$ with the polynomial in $x$ of degree $n-1$
which coincides with $f(x)$ at the $n$ points.

Coefficients up to the six-point formula are given in table 25.2
of Ref.~\cite{AS70}.
We have utilized a symbolic computation software
{\sl mathematica}\cite{Wol91}
to obtain the values of the coefficients of higher-order formulae,
which are given in Table~\ref{t_numderiv}.

\begin{table}[hbt]
\caption{Coefficients of the $n$-point approximation formulae
 for the $\nu$th derivative given in Eq.~(\protect\ref{eq:numderiv}).
 $c_{ni}^{(\nu)}$ for negative $i$ can be obtained using a relation
 $c_{n,-i}^{(\nu)}=(-1)^{\nu} c_{ni}^{(\nu)}$.
 $c_{n0}^{(\nu)}=0$ for odd $\nu$.
 \label{t_numderiv} }
\begin{center}
Sign of $c_{ni}^{(\nu)}$
\begin{tabular}{|c|c|c|c|}
\hline
 & $i<0$ & $i=0$ & $i>0$ \\
\hline
$\nu =1$ & $(-1)^i$ & 0 & $(-1)^{i+1}$ \\
\hline
$\nu =2$ & \multicolumn{3}{c|}{$(-1)^{i+1}$} \\
\hline
\end{tabular}
\\
\vspace*{3mm}
First derivative ($\nu=1$)
\begin{tabular}{|r|r|l|}
\hline
$n$ & $b_n^{(1)}$ & $|c_{ni}^{(1)}|$; $i=1,2,\cdots,\frac{1}{2}(n-1)$\\
\hline
 3 & 2 &  1\\
 5 & 12 &  8, 1\\
 7 & 60 &  45, 9, 1\\
 9 & 840 &  672, 168, 32, 3\\
 11 & 2520 &  2100, 600, 150, 25, 2\\
 13 & 27720 &  23760, 7425, 2200, 495, 72, 5\\
 15 & 360360 &  315315, 105105, 35035, 9555, 1911, 245, 15\\
 17 & 720720 &  640640, 224224, 81536, 25480, 6272, 1120, 128, 7\\
 19 & 12252240 &  11027016, 4009824, 1559376, 539784, 154224,
                 34272, 5508, 567, 28\\
 21 & 232792560 &  211629600, 79361100, 32558400, 12209400, 3907008,
                  1017450, 205200,\\
    &           & 29925, 2800, 126\\
\hline
\end{tabular}
\\
\vspace*{3mm}
Second derivative ($\nu=2$)
\begin{tabular}{|r|r|l|}
\hline
$n$ & $b_n^{(2)}$ & $|c_{ni}^{(2)}|$; $i=0,1,2,\cdots,\frac{1}{2}(n-1)$\\
\hline
 3 & 1 & 2, 1\\
 5 & 12 & 30, 16, 1\\
 7 & 180 & 490, 270, 27, 2\\
 9 & 5040 & 14350, 8064, 1008, 128, 9\\
 11 & 25200 & 73766, 42000, 6000, 1000, 125, 8\\
 13 & 831600 & 2480478, 1425600, 222750, 44000, 7425, 864, 50\\
 15 & 75675600 & 228812298, 132432300, 22072050, 4904900, 1003275,
                  160524, 17150, 900 \\
 17 & 302702400 & 924708642, 538137600, 94174080, 22830080, 5350800,
                   1053696, 156800,\\
    &           &  15360, 735\\
 19 & 15437822400 & 47541321542, 27788080320, 5052378240, 1309875840,
                    340063920,\\
    &             & 77728896, 14394240, 1982880, 178605, 7840\\
 21 & 293318625600 & 909151481810, 533306592000, 99994986000,
                     27349056000, 7691922000,\\
    &              & 1969132032, 427329000, 73872000, 9426375,
                     784000, 31752\\
\hline
\end{tabular}
\end{center}
\end{table}

\begin{figure}
\centerline{\includegraphics[angle=-90, width=135mm]{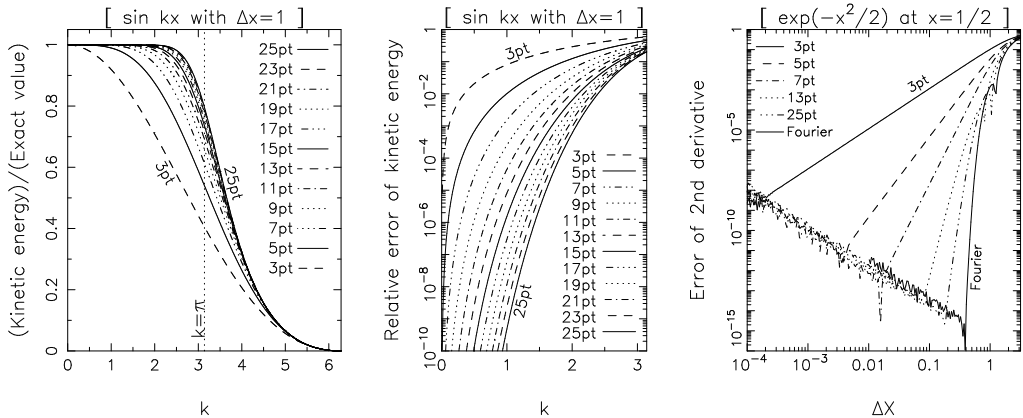}}
\caption{
Precision of finite-point numerical differentiation formulae.
} \label{g_numdrv}
\end{figure}

We show in Fig.~\ref{g_numdrv} the results of
our calculations with which we examine the precision of the kinetic energy
calculated using Eq.~(\ref{eq:numderiv}).
The results are for one-dimensional case but can be easily applied
to three-dimensional cases because the kinetic energy
is divided into $x$, $y$ and $z$ components.

In the left-hand portion of Fig.~\ref{g_numdrv}, the ratio of the
numerically calculated value of
$\int_{0}^{2\pi/k} \sin kx \frac{d^2}{dx^2} \sin kx dx$
to its exact value $-k\pi$ is plotted as a function of the
wave number $k$.
The mesh spacing $a$ (expressed as $\Delta x$ in the figure) is taken
as 1 fm.
We change $k$ from 0 to $2 \pi$ for an aesthetic reason, though $k >
\pi/a = \pi \; {\rm fm}^{-1}$
is meaningless from the point of view of the Fourier expansion.
In the numerical calculation, the second derivative of $\sin kx$
is approximated by that of the $(2n+1)$-point Lagrange polynomial,
which has degree $n$ and coincides with $\sin kx$ at
$x, x\pm a, \cdots, x\pm n a$.
The number of points is chosen as $n = 3,5,\cdots,25$.
The integral, too, is evaluated numerically but with much higher
precision than that of the numerical differentiation.
Hence the plotted quantity is the error of numerical differentiation,
not that of integration.
%
%
%
From this result one can see that the value of kinetic energy is always
underestimated and the error becomes smaller by employing higher-order
formulae.
%
%

The middle portion of Fig.~\ref{g_numdrv} is the same as the left-hand
portion except that the ordinate is the relative error $|
\mbox{(ratio)} - 1 |$ in logarithmic scale.
One can read off from the figure, for example, that the relative error
of the kinetic energy at $k=1.4$ fm$^{-1}$ is $3.0 \times 10^{-3}$ for
the 9-point formula of the second derivative. The 9-point formula is
used in the calculations presented in this paper.  This error roughly
accounts for the error of the total energy of $^{240}$Pu, which is
0.5\%.
One can conclude that the main source of the error of the total energy
is the kinetic energy.

In the right-hand portion of Fig.~\ref{g_numdrv}, the error of numerically
evaluated values of the second derivative of a function
$e^{-x^2/2}$ at $x=\frac{1}{2}$ is shown as a
function of the mesh spacing $a$ (called $\Delta x$ in the figure).
The numerical evaluation means the usage of
the $(2n+1)$-point Lagrange polynomial which coincides with
the function $e^{-x^2/2}$
at $x=\frac{1}{2},\frac{1}{2}\pm a, \cdots ,\frac{1}{2}\pm n a$.
The solid curve labeled as ``Fourier'' indicates the error of the
Fourier interpolation, which coincides with the original function
$e^{-x^2/2}$ at infinite number of discrete points of $x=\frac{1}{2}+
n a \; (n = 0, \pm 1, \pm 2, \cdots$).  For plane waves with $k <
\pi/a$, the Fourier interpolation gives the exact value. For
functions involving higher wave-number components, as the present
Gaussian function, the Fourier method does not result in the exact
value. Nevertheless,
as one can see from the figure, it is still much more
precise than polynomial-based formulae.

In order to evaluate the derivative values at points within
$\frac{1}{2}(n-1)a$ from the boundaries, one needs the values of the
function beyond the boundaries. In order to obtain them one has to
assume either vanishing (reflection antisymmetric) or periodic
boundary conditions.  A much easier method is to assume the function
to be zero beyond the boundaries. However it is justified only when
the function is well-damped before reaching the boundaries.  Otherwise
it produces a discontinuity of derivative values at the boundaries and
invalidate the $n$-point formulae for $n \ge 3$.
Usage of a cavity whose shape is not a box
(rectangular parallelepiped)
can be justified only for
functions well-localized around the center of the cavity.


\section{Finite-Point Formulae for Numerical Integration}

The integral of a function $f(x)$ over an interval $0 \le x \le L$ can
be evaluated by the values of the function at $n+1$ mesh points, $x_i
= a i$ ($0 \le i \le n$, $a=L/n$), by approximating the function with
a polynomial in $x$ of degree $n$ coinciding with $f(x)$ at the
mesh points. Such formulae are called ($n+1$)-point
Newton-Cotes closed type integration formula and are expressed as
\begin{equation} \label{e_NC_integ}
\int_0^{na} f(x) dx = a \sum_{i=0}^{n} \frac{\alpha_{ni}}{\beta_n} f(ai)
+{\cal O}(a^{n+\delta}),
\;\;\; \delta=\left\{\begin{array}{l} \! 3 \;\;\; (\mbox{even} \; n)\\
\! 2 \;\;\; (\mbox{odd} \; n) \end{array}\right. .
\end{equation}
We have
calculated the values of the coefficients $\alpha_{ni}$ and $\beta_{n}$
in a similar method as we calculated those of the differentiation
formulae.  The results are given in Table~\ref{t_NC_integ} for some
odd values of $n+1$.

\begin{table}[hbt]
\caption{Coefficients of Newton-Cotes closed type $(n+1)$-point
integration formulae.
Values for $i>\frac{1}{2}(n+1)$ can be obtained from a relation
$\alpha_{ni} = \alpha_{n,n-i}$.
\label{t_NC_integ} }
\begin{center}
\begin{tabular}{|r|r|l|}
\hline
$n+1$ & $\beta_n$ & $\alpha_{ni}$; $i=0,1,\cdots,\frac{1}{2}n+1$\\
\hline
3 & 3 & 1, 4 \\
5 &  45 & 14, 64, 24 \\
7 & 140 & 41, 216, 27, 272 \\
9 & 14175 & 3956, 23552, -3712, 41984, -18160 \\
11 & 299376 & 80335, 531500, -242625, 1362000, -1302750, 2136840 \\
13 &  5255250 & 1364651, 9903168, -7587864, 35725120, -51491295, 87516288,
      -87797136 \\
15 & 2501928000 & 631693279, 4976908048, -5395044599, 24510099488,\\
 & &  -46375653541, 88410851312, -117615892611, 136741069248 \\
\hline
\end{tabular}
\end{center}
\end{table}
%
Two-, three-, and four-point formulae are called respectively
the trapezoidal rule, Simpson's $\frac{1}{3}$ rule,
and Simpson's $\frac{3}{8}$ rule.
The errors of $2k$-point and ($2k-1$)-point formulae are of
the same order in $a$, ${\cal O}(a^{2k+1})$, for $2k \ge 4$.
For this reason we listed in Table~\ref{t_NC_integ}
only 2- and odd-point formulae.
The coefficients of 4- to 11-point formulae agree with
Eqs.~25.4.13 -- 20 of Ref.~\cite{AS70}.
There is no difficulty
to obtain the coefficients of formulae which involve
more number of points.
It should be kept in mind, however, that
formulae for $n+1 \ge =9$ ($n+1 =10$ is an exception) include negative
coefficients, which cause a loss of numerical precision due to
cancellations between the contributions of
negative and positive coefficient points.  This
problem becomes more serious for larger $n$.

In order to apply the ($n+1$)-point formula to a long interval,
one divides
the interval into $m$ segments and
applies the $(n+1)$-point formula to each of the segments, i.e.,
\begin{equation} \label{e_eNC_integ}
\int_0^{mna} f(x) dx = a \sum_{j=0}^{m-1} \sum_{i=0}^{n}
\frac{\alpha_{ni}}{\beta_n} f \left( (i+nj) a \right)
+{\cal O}(a^{n+\delta'}),
\;\;\; \delta'=\left\{\begin{array}{l} \! 2 \;\;\; (\mbox{even} \; n)\\
\! 1 \;\;\; (\mbox{odd} \; n) \end{array}\right. .
\end{equation}
Note that points $x=jna$ ($1 \le j \le m-1$) appear twice in the
double summation.
The reason for $\delta'=\delta-1$ is that
the error of the integral over the interval length $L=mna$ is
$m = L/na \propto a^{-1}$ times as large as
the error on the right-hand side on Eq.~(\ref{e_NC_integ}).

\begin{figure}
\centerline{\includegraphics[angle=-90, width=135mm]{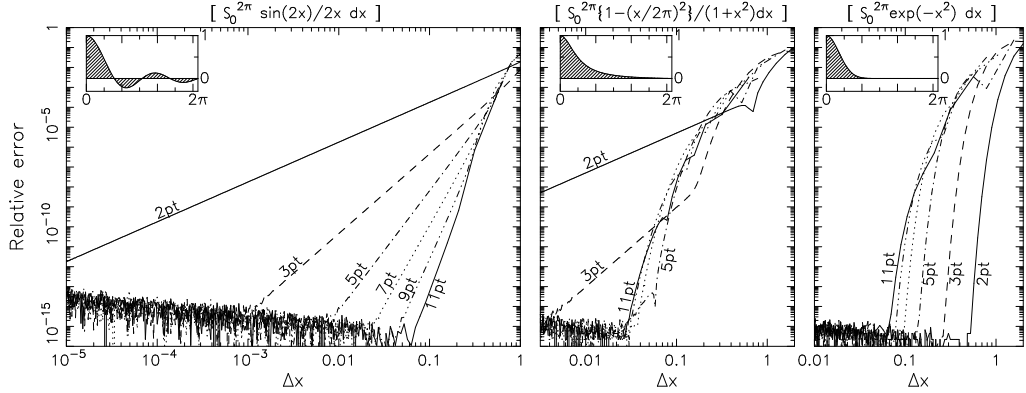}}
\caption{
Errors of numerical integral with Newton-Cotes closed type formulae.
See text for explanations.
} \label{g_numint}
\end{figure}

We have calculated the precision of these formulae as a function of the
mesh spacing $a$ for three types of integrals,
\begin{equation} \label{e_integrands}
\int_{0}^{2\pi} \frac{\sin 2x}{2x} dx,          \;\;\;\;\;\;\;
\int_{0}^{2\pi} \frac{1- \left(\frac{x}{2\pi}\right)^2}{1 + x^2} dx,
\;\;\;\;\;\;\;
\int_{0}^{2\pi} e^{-x^2} dx,
\end{equation}
as examples of three different boundary situations at $x=2\pi$.  The
obtained (absolute value of the) relative errors of the integrals are
shown in logarithmic scale versus the mesh spacing $a$ ($= \Delta x$)
in Fig.~\ref{g_numint}.
The calculations are defined only for discrete points at $a =
2\pi/mn$. The lines drawn in the figure connect such discrete points.

In the left-hand portion of Fig.~\ref{g_numint},
the error of the $n$-point formula behaves as expected
from Eq.~(\ref{e_eNC_integ}) for every value of $n$.
More-point formulae give quicker decreases of the error as the mesh
spacing decreases.
The noise-like error of magnitude $10^{-13}$ to $10^{-15}$ originates
in the round off of real numbers into double precision (16-digit)
floating numbers.

However, this is not the only situation.
We now show that the convergence to the exact value is much faster for
some kind of integrands and, what is more interesting is, that such
fast convergences become more outstanding by using low-order few-point
formulae than high-order many-point ones.

%
%

The right-hand portion of Fig.~\ref{g_numint} shows the results for a
different integrand $e^{-x^2}$.  In this case, one can see that
fewer-point formulae give more accurate results than more-point
formulae.  The convergences are generally much faster than those in
the left-hand portion of the figure.

The origin of this unexpected situation
apparently in contradiction with Eq.~(\ref{e_eNC_integ})
is the approximate periodicity
of the integrand.  The function $e^{-x^2}$ in an interval $[0, 2\pi]$
can be regarded as a half wave length portion of a periodic function,
which is constructed using this portion of the function
by repeating reflections at the boundaries $x=0$ and $2\pi$.
%
%
At $x=0$ the function is reflection symmetric originally.
For $x>2\pi$ the function is redefined by a reflection symmetry
requirement $f(2\pi + x) = f(2\pi -x)$.  This reflection makes
practically no discontinuities at $x=2\pi$ (or between $x=\pm 2 \pi$
because the period is $4\pi$) because $e^{-x^2}$ and its derivatives
(of not extremely high order;
the graph includes up to 11-point formula
and thus only up to 11th derivative concerns) are close to zero at
$x=2\pi$.
%
%

The integral over $[0,2\pi]$ is a half of the integral of thus
constructed smooth periodic function over the wave length $4\pi$.  It
is known that the best numerical formula for integrals of periodic
functions over a wave length is the mid-point rule.  The reason is
often explained in terms of an asymptotic expansion.  A more
quantitatively accurate explanation can be given in terms of Fourier
expansions, as explained in Appendix \ref{a_fourierbasis}.

Periodic functions composed of wave numbers less than $\pi/a$
can be exactly integrated with the mid-point rule.  The function
$e^{-x^2}$ is an approximately periodic function and its
large-wave-number components are non-zero but very small.
Therefore the mid-point rule is the most precise formula for this
integrand.  The 2-point formula is the same as the mid-point rule in
the present situation

The $(n+1)$-point formulae can be regarded as a summation of $n$
similar results of 2-point formula
with mesh spacing of $n a$.  The curve of
the error of the 3-point formula resembles to that of the
2-point formula except it is shifted leftward by a factor $(2-1)/(3-1)
= \frac{1}{2}$.  Similarly, the horizontal distance between
the curves for 3- and
5-point formulae is also calculated as $(3-1)/(5-1)=\frac{1}{2}$.

We have searched for an integrand which gives rise to an intermediate
situation between the left-hand and the right-hand portions of
Fig.~\ref{g_numint}.
A function which serves for this aim is
$ \left\{ 1- (x/2\pi)^2 \right\}/ \left( 1 + x^2 \right)$.
The middle portion of Fig.~\ref{g_numint} shows the error of the
integral for this integrand.
One can see that the situation is like that of the left-hand (right-hand)
portion of the figure for small (large) $a$ ($= \Delta x$).
The most precise formula changes with the mesh spacing:
It is the 2-point formula for $a > 0.3$,
the 3-point formula for $0.08 < a < 0.3$,
and the 5-point formula for $a < 0.08$.

This change can be understood as follows.
With large $a$,
the precision is so low that
the discontinuity at $x=2\pi$ looks too small to
invalidate the approximation of the periodicity.
Consequently, the error behaves like a parabola
as in the right-hand portion of the figure.
With small $a$, the precision is so high that
the discontinuity looks large enough to violate the periodicity.
Thus the error behaves linearly
as in the left-hand portion of the figure.
%
%
%

In the nuclear mean-field calculations, the integrands are
not exactly periodic.  However, the situation resembles that of the
right-hand portion of Fig.~\ref{g_numint} for nuclei whose Fermi levels
are not very shallow, because the wave functions are well damped
before reaching to the boundaries. This is why we employ the mid-point
rule in our calculations.


\begin{table}[hbt]
\caption{Coefficients of Boundary-correction type $(n+1)$-point
integration formulae.}
\label{t_BC_integ}
\begin{center}
\begin{tabular}{|r|r|l|}
\hline
$n+1$ & $\beta'_n$ & $\alpha'_{Nni}$; $i=0,1,\cdots,n$\\
\hline
 2 &  2 &
 1, 2 \\
 3 & 24 &
 9, 28, 23 \\
 4 & 24 &
 8, 31, 20, 25 \\
 5 & 1440 &
 475, 1902, 1104, 1586, 1413 \\
 6 & 1440 &
 459, 1982, 944, 1746, 1333, 1456 \\
 7 & 120960 &
 36799, 176648, 54851, 177984, 89437, 130936, 119585 \\
 8 & 120960 &
 35584, 185153, 29336, 220509, 46912, 156451, 111080, 122175 \\
 9 & 7257600 &
 2082753, 11532470, 261166, 16263486, -1020160, 12489922, 5095890,\\ & &
 7783754, 7200319 \\
 10 & 7257600 &
 2034625, 11965622, -1471442, 20306238, -7084288, 18554050, 1053138,\\ & &
 9516362,  6767167, 7305728 \\
 11 & 958003200 &
 262747265, 1637546484, -454944189, 3373884696, -2145575886, \\ & &
 3897945600, -1065220914, 1942518504, 636547389, 1021256716, \\ & &
 952327935 \\
 12 & 958003200 &
 257696640, 1693103359, -732728564, 4207237821, -3812282136,\\ & &
 6231334350, -3398609664, 3609224754, -196805736, 1299041091,\\ & &
 896771060, 963053825 \\
 13 & 5230697472000 &
 1382741929621, 9535909891802, -5605325192308, 28323664941310,\\ & &
 -32865015189975, 53315213499588, -41078125154304, 39022895874876,\\ & &
 -13155015007785, 12465244770050, 3283609164916, 5551687979302,\\ & &
 5206230892907 \\
 14 & 5230697472000 &
 1360737653653, 9821965479386, -7321658717812, 34616887868158,\\ & &
 -48598072507095, 81634716670404, -78837462715392, 76782233435964,\\ & &
 -41474518178601, 28198302087170, -3009613761932, 7268021504806,\\ & &
 4920175305323, 5252701747968 \\
 15 & 62768369664000 &
 16088129229375, 121233187986448, -109758975737401, \\ & &
  502985565327936, -823993097730133, 1461175500619600,\\ & &
  -1668277571373981, 1746664157478912, -1219701572786787, \\ & &
  819644306759856, -276710928642475, 174692180291008,\\ & &
   37176466501689, 66395850785776, 62528161418177 \\
\hline
\end{tabular}
\end{center}
\end{table}
%

There are also modified types of Newton-Cotes formulae, in which the
coefficients are different from 1 only near the boundaries:
\begin{equation} \label{e_BC_integ}
\int_0^{Na} f(x) dx = a \sum_{i=0}^{N}
\frac{\alpha'_{Nni}}{\beta'_n} f \left( ai \right)
+{\cal O}(a^{n+\delta'}),
\;\;\; \delta'=\left\{\begin{array}{l} \! 2 \;\;\; (\mbox{even} \; n)\\
\! 1 \;\;\; (\mbox{odd} \; n) \end{array}\right. ,
\end{equation}
with
\begin{eqnarray}
\alpha'_{N,n,i} \; = & \;\alpha'_{N,n,N-i} \;\;\; & ( 0 \le i \le n ),
\label{e_BC_integ2}\\
\alpha'_{Nni} \; = & \beta'_n \;\;\; & ( n+1 \le i \le N-n-1 ).
\label{e_BC_integ3}
\end{eqnarray}
We temporarily call the above equations the boundary-correction type
formulae in this paper.
We have calculated the values of the coefficients $\alpha'_{Nni}$ and
$\beta'_{n}$ in a similar manner as we treated Eq.~(\ref{e_NC_integ}).
Table~\ref{t_BC_integ} gives the values of the coefficients.
For ($n+1$)-point formula, the coefficients are different from one
only for $n+1$ points near each ends of the integration interval.
Only in the 2-point formula, which is identical to the
extended trapezoidal rule, the end points alone have a non-unit
coefficient.
We give both $2k$-point and ($2k-1$)-point formulae in
the table.
Although their errors are of the same order in $a$,
the former seems slightly more precise than the latter
in some test calculations.
We have found only the 3-point formula in literature.
It is Eq.~(4.1.14) of Ref.~\cite{PTV92}.
Although the formulae are identical, the derivation is quite different.

Boundary-correction type formulae will be useful to treat
deeply bound, weakly bound, and positive-energy orbitals in a single
framework.  With the formulae, the deeply bound orbitals are treated
with the high precision of the mid-point rule for periodic integrands
while the weakly bound and positive-energy orbitals are treated with
less precise but still high-precision $(n+1)$-point formula.

%
%


\vspace*{\baselineskip}
\fbox{
\fbox{
\begin{minipage}[t]{130mm}
This paper has been published in
Progress of Theoretical Physics
Supplement No.~142 (2001) on Physics of Unstable Nuclei, pp. 265-296.\\
The author converted the souce file of the paper 
into Latex2e format in order to 
inclde PostScript figures and then uploaded it to 
the e-print archive {\em arXiv} on July 22, 2003.
\end{minipage}
}
}


\begin{thebibliography}{99}
\bibitem{We35}  
         C.F.\ von Weizs{\"a}cker, Z.\ Phys.\ {\bf 96} (1935) 431.
\bibitem{BB36}  
  H.A.\ Bethe and R.F.\ Bacher, Rev.\ Mod.\ Phys.\ {\bf 8} (1936) 82.
\bibitem{RS80}
         P. Ring and P. Schuck, {\em The nuclear many-body problem}
         (Springer, New York, 1980)
\bibitem{AW93}  
          G.~Audi and A.H.~Wapstra, Nucl.Phys. {\bf A565} (1993) 1.
\bibitem{AW95}  
          G.~Audi and A.H.~Wapstra, Nucl.Phys. {\bf A595} (1995) 409;\\
          http://www.nndc.bni.gov/nndcscr/masses/
\bibitem{TUY88} 
         T.\ Tachibana, M.\ Uno, M.\ Yamada, and S.\ Yamada,
         Atomic Data Nucl. Data Tables {\bf 39} (1988) 251.
\bibitem{KY00} 
        H.~Koura and M.~Yamada, Nucl.\ Phys.\ {\bf A671} (2000) 96.
\bibitem{KUT00} 
        H.~Koura, M.~Uno, T.~Tachibana, and M.~Yamada,
        Nucl.\ Phys.\ {\bf A674} (2000) 47.
\bibitem{DZ95} 
         J.\ Duflo and A.P.\ Zuker, Phys.\ Rev.\ {\bf C52} (1995) R23.
\bibitem{Uno00} 
        M.~Uno, proc. Models and Theories of the Nuclear Mass,
        RIKEN Review {\bf 26} (2000) 38.
\bibitem{MNM95} 
         P.~M{\"o}ller, J.R.~Nix, W.D.~Myers, and W.J.~Swiatecki,
         Atomic Data Nucl. Data Tables {\bf 59} (1995) 185.
\bibitem{APD92} 
         Y.~Aboussir, J.M.~Pearson, A.K.~Dutta, and F.~Tondeur,
         Nucl.\ Phys.\ {\bf A549} (1992) 155;
         Atomic Data Nucl. Data Tables {\bf 61} (1995) 127.
\bibitem{TTO96} 
         N.~Tajima, S.~Takahara, and N.~Onishi,
         Nucl.\ Phys.\ {\bf A603} (1996) 23-49;\\
         http://serv.apphy.fukui-u.ac.jp/$\sim$tajima/hfs3/index.html
         \hspace*{2mm}.
\bibitem{TGP00} 
        F.~Tondeur, S.~Goriely, J.M.~Pearson, and M.~Onsi,
        Phys.\ Rev.\ {\bf C62} (2000) 024308.
\bibitem{GTP01} 
        S.~Goriely, F.~Tondeur, and J.M.~Pearson,
        to appear in Atomic Data and Nuclear Data Tables.
\bibitem{SHT96} 
        K.~Sumiyoshi, D.~Hirata, I.~Tanihata, Y.~Sugahara, and H.~Toki,
        RIKEN Review {\bf 14} (1996) 25.
\bibitem{Va73}  
         D.~Vautherin, Phys.Rev.{\bf C7} (1973) 296.
\bibitem{DG80}  
         J.~Decharg\'{e} and D.~Gogny,
         Phys. Rev. {\bf C21} (1980) 1568.
\bibitem{BFH85} 
         P.~Bonche, H.~Flocard, P.-H.~Heenen, S.J.~Krieger, and
         M.S.~Weiss, Nucl.\ Phys.\ {\bf A443} (1985) 39.
\bibitem{Tan85} 
         I.~Tanihata et al., Phys.\ Rev.\ Lett.\ {\bf 55} (1985) 276.
\bibitem{TOT94} 
         N.~Tajima, N.~Onishi, S.~Takahara,
         Nucl.\ Phys.\ {\bf A588} (1995) 215c-220c.
\bibitem{TTO98} 
         S.~Takahara, N.~Tajima, and N.~Onishi,
         Nucl.\ Phys.\ {\bf A642} (1998) 461-479.
\bibitem{Sk56}  
         T.H.R.~Skyrme, Phil.\ Mag.\ {\bf 1} (1956) 1043.
\bibitem{VB72}  
         D.~Vautherin and D.M.~Brink,
         Phys.\ Rev.\ {\bf C5} (1972) 626.
\bibitem{TBF93} 
         N. Tajima, P. Bonche, H. Flocard, P.-H. Heenen, and
         M.S. Weiss, Nucl. Phys. {\bf A551} (1993) 434.
\bibitem{BFG75} 
         M. Beiner, H. Flocard, Nguyen van Giai, and P. Quentin,
         Nucl.Phys. {\bf A238} (1975) 29.
\bibitem{Ham62}
  M.~Hamermesh, {\em Group theory and its application to physical problems}
  ( Addison-Wesley, 1964), chap.~2-7.
\bibitem{Taj98} 
  N. Tajima, proc. Innovative Computational Methods in Nuclear Many-Body
  Problems, Osaka, Japan, 1997, edited by H.~Horiuchi et al.,
  (1998) World Scientific (Singapore) , p.~343.
\bibitem{Taj00a} 
        N.~Tajima, proc. Models and Theories of the Nuclear Mass,
        RIKEN Review {\bf 26} (2000) 87.
\bibitem{RMM87} 
         S.\ Raman, C.H.\ Malarkey, W.T.\ Milner, C.W. Nestor, JR.,
         and P.H.\ Stelson,
         Atomic Data Nucl.\ Data Tables {\bf 36} (1987) 1.
\bibitem{RSB95}
         S.\ Raman, J.A.\ Sheikh, and K.H.\ Bhatt,
         Phys.\ Rev.\ {\bf C52} (1995), 1380.
\bibitem{BQB82} 
         J. Bartel, P. Quentin, M. Brack, C. Guet, and
         H.-B. Hakansson, Nucl.Phys. {\bf A386} (1982) 79.
\bibitem{GS81} 
         Nguyen Van Giai and
         H. Sagawa, Phys. Lett {\bf B106}, 379 (1981).
\bibitem{Vo74}  
         A.B.~Volkov, Nucl.\ Phys.\ {\bf 74} (1965) 33.
\bibitem{LL75}  
         G.~Leander, S.E.~Larsson, Nucl.\ Phys.\ {\bf A239} (1975) 93.
\bibitem{KH95}  
         Y.~Kanada-En'yo and H.~Horiuchi,
         Prog.\ Theor.\ Phys.\ {\bf 93} (1995) 115.
\bibitem{ZZC91} 
         L.~Zamick, D.C.~Zheng, J.A.~Caballero, E. Moya de Guerra,
         Ann.\ Phys.\ {\bf 212} (1991) 402.
\bibitem{TS01} 
        Naoki Tajima and Norifumi Suzuki, nucl-th/0103061.
\bibitem{FOT93} 
         N.~Fukunishi, T.~Otsuka, and I.~Tanihata,
         Phys.\ Rev.\ {\bf C48} (1993) 1648.
\bibitem{BH86}  
         D. Baye and P.-H. Heenen, J. Phys. {\bf A19} (1986) 2041.
\bibitem{AS70} %
         Handbook of mathematical functions, 9th printing,
         ed. M.~Abramowitz and I.A.~Stegan, Dover, N.Y. (1970).
\bibitem{Wol91} %
   S.\ Wolfram, {\em Mathematica, a System for Doing Mathematics by Computer}
   (Addison-Wesley, 1991).
\bibitem{PTV92} 
         W.H.~Press, S.A.~Teukolsky, W.T.~Vetterling, and B.P.~Flannery
         {\em Numerical Recipes in C} (Cambridge Univ.\ Press, 1992).
\end{thebibliography}
\end{document}